\def\PaperFullBuild{}
\documentclass[acmsmall,screen]{acmart}

\makeatletter
\@ifpackageloaded{unicode-math}{\DeclareSymbolFont{symbols}{OMS}{cmsy}{m}{n}
  \DeclareSymbolFont{largesymbols}{OMX}{cmex}{m}{n}
}{}
\makeatother

\usepackage{cleveref}
\usepackage{mathpartir}
\usepackage[normalem]{ulem}
\usepackage{stmaryrd}

\makeatletter
\@ifpackageloaded{unicode-math}{
  
  \let\digamma\relax
  
}{}
\makeatother
\usepackage{amssymb}
\usepackage{hyperref}
\usepackage{subcaption}
\usepackage{enumitem}
\usepackage{pdflscape}

\usepackage{tikz}
\usetikzlibrary{arrows.meta}
\usetikzlibrary{positioning}
\usetikzlibrary{intersections}
\usetikzlibrary{shapes.geometric}
\usetikzlibrary{shapes.multipart}
\usetikzlibrary{calc}

\usepackage{listings, listings-rust}
\usepackage{wrapfig}

\definecolor{Cerulean}{rgb}{0.20,0.31,0.0.90}
\definecolor{RedOrange}{rgb}{0.70,0.59,0.0.20}
\definecolor{ForestGreen}{rgb}{0.1,0.60,0.2}
\definecolor{SomeRed}{rgb}{0.80,0.3,0.0.20}
\definecolor{lightGray}{rgb}{0.85,0.85,0.86} \definecolor{lGray}{rgb}{0.31,0.31,0.33} \definecolor{lYellow}{rgb}{0.99,0.78,0.07} \definecolor{lila}{rgb}{0.7,0.30,0.77} \definecolor{cloudyblue}{rgb}{206,221,246}

 \usepackage{ifthen}

\newboolean{showcomments}
\setboolean{showcomments}{false}

\makeatletter
\newcommand{\mynote}[3]{\ifthenelse{\boolean{showcomments}}{\fbox{\bfseries\sffamily\scriptsize#1}{\small\textsf{\emph{\color{#3}{#2}}}}}{\@bsphack
   \@esphack
  }}
\makeatother

\definecolor{asparagus}{rgb}{0.53, 0.66, 0.42}
\newcommand{\todo}[1]{\mynote{TODO}{#1}{red}}

\newcommand{\se}[1]{\mynote{Sebastian}{#1}{brown}}
\newcommand{\ff}[1]{\mynote{Franta}{#1}{violet}}

\newcommand{\crmn}[1]{\mynote{Carmine}{#1}{blue}}

\newcommand{\out}[1]{}

\setboolean{showcomments}{false}
 \usepackage{xspace}

\newcommand{\itree}{\text{ITree}\xspace}
\newcommand{\itrees}{\text{ITrees}\xspace}

\newcommand{\iProp}{\mathit{iProp}} \newcommand{\updateMod}{\mid \! \Rrightarrow}

\newcommand*{\lstmath}[1]{{\text{\lstinline{#1}}}}

\newcommand*{\wand}[0]{\text{-}\!\!\ast}
\newcommand{\locT}[0]{\texttt{loc}}

\newcommand{\lobind}[0]{L{\"o}b}
\newcommand{\itreeER}[2]{\ensuremath{\texttt{itree}\;{#1}\;{#2}}}
\newcommand{\Ret}[1]{\ensuremath{\texttt{Ret}(#1)}}
\newcommand{\Tau}[1]{\ensuremath{\texttt{Tau}(#1)}}
\newcommand{\VisA}[3]{\ensuremath{\texttt{Vis}_{#1}(#2, #3)}}
\newcommand{\Vis}[2]{\ensuremath{\texttt{Vis}(#1, #2)}}
\newcommand{\trigger}[1]{\ensuremath{\texttt{trigger}\;#1 }}
\newcommand{\itreeof}[1]{\ensuremath{ \llbracket #1 \rrbracket }}
\newcommand{\heap}[0]{\texttt{heap}}
\newcommand{\stateE}[1]{{\ensuremath{\texttt{StateE}_{#1}}}}
\newcommand{\getE}[0]{\texttt{getE}}
\newcommand{\setE}[1]{\ensuremath{\texttt{setE}(#1)}}
\newcommand{\ubE}[0]{\ensuremath{\texttt{ubE}}}
\newcommand{\stepE}[0]{\ensuremath{\texttt{stepE}}}
\newcommand{\thirE}[0]{\ensuremath{\texttt{thirE}}}
\newcommand{\store}[2]{\ensuremath{\texttt{store}\;#1\;#2}}
\newcommand{\storeorub}[2]{\ensuremath{\texttt{store\_or\_ub}\;#1\;#2}}
\newcommand{\wpi}[3]{\ensuremath{ \mathrm{wpi}_{#1}\;#2\;\{ #3 \} }}
 \newcommand{\sys}[1]{Corten}

\newcommand{\valBool}[1]{\mathtt{bool}~#1}
\newcommand{\valPtr}[1]{\mathtt{ptr}~#1}
\newcommand{\valInt}[3]{\mathtt{int}~#1~#2~#3}
\newcommand{\valTuple}[1]{\mathtt{tuple}~#1}
\newcommand{\valUnit}[1]{\mathtt{unit}~#1}
\newcommand{\valAdt}[2]{\mathtt{adt}~#1~#2}
\newcommand{\valStruct}[2]{\mathtt{struct}~#1~#2}
\newcommand{\valEnum}[3]{\mathtt{enum}~#1~#2~#3}

\newcommand{\lookupTuple}[2]{#1~!!_\mathtt{tuple}~#2}
\newcommand{\lookupAdt}[2]{#1~!!_\mathtt{adt}~#2}

\newcommand{\RTyBool}{\mathtt{bool}_\text{R}}
\newcommand{\RTyPtr}{\mathtt{ptr}_\text{R}}
\newcommand{\RTyInt}[2]{\mathtt{int}_\text{R}}
\newcommand{\RTyTuple}[1]{\mathtt{tuple}_\text{R}~#1}

\newcommand{\RTyStruct}[1]{\mathtt{struct}_\text{R}~#1}
\newcommand{\RTyEnum}[1]{\mathtt{enum}_\text{R}~#1}

\newcommand{\toRTy}[1]{[\![ #1 ]\!]_{\RTy}}

\newcommand{\locBase}[1]{\mathtt{base}~#1}
\newcommand{\locField}[2]{#1.#2}
\newcommand{\locIx}[2]{#1.#2}
\newcommand{\locConst}[1]{\mathtt{const}~#1}

\newcommand{\Type}{\mathit{Type}}

\newcommand{\Pat}{\mathit{Pat}}
\newcommand{\Stmt}{\mathit{Stmt}}
\newcommand{\Expr}{\mathit{Expr}}
\newcommand{\Loc}{\mathcal{L}}
\newcommand{\Val}{\mathcal{V}}

\newcommand{\Spec}{\mathit{Spec}}
\newcommand{\RTy}{\mathit{Type}_\text{R}}

\definecolor{syntaxblue}{RGB}{53, 125, 183}
\lstdefinestyle{inlinestyle}{
  basicstyle=\ttfamily
    \color{syntaxblue}\lst@ifdisplaystyle\footnotesize\fi,emph={identity}
}

\renewcommand*{\lstmath}[1]{{\text{\lstinline[style=inlinestyle]{#1}}}}

\newcommand{\Rbool}{\lstmath{bool}}
\newcommand{\Rchar}{\lstmath{char}}
\DeclareMathOperator*{\Rnumaux}{\lstmath{num}}
\newcommand{\Rnum}[2]{\Rnumaux #1~#2}
\newcommand{\Rtuple}[1]{\lstmath{(}#1\lstmath{)}}
\newcommand{\Rfun}[2]{\lstmath{fn(}#1\lstmath{)} \rightarrow~#2}
\newcommand{\Rarray}[2]{\lstmath{[}#1 \lstmath{;} #2\lstmath{]}}
\newcommand{\Rslice}[1]{\lstmath{[}#1\lstmath{]}}
\newcommand{\Runit}{\lstmath{()}}

\DeclareMathOperator*{\Radtaux}{\lstmath{adt}}
\newcommand{\Radt}[1]{\Radtaux #1}
\newcommand{\Rrefsym}{\lstmath{&}}
\newcommand{\Rref}[1]{\Rrefsym #1}
\newcommand{\Rrefmutsym}{\lstmath{&mut}}
\newcommand{\Rrefmut}[1]{\Rrefmutsym~#1}

\newcommand{\Rnever}{\lstmath{!}}

\newcommand{\Rpatwild}{\lstmath{_}}
\newcommand{\Rderef}{\lstmath{*}}
\newcommand{\Rcast}[2]{#1~\lstmath{as}~#2}
\newcommand{\Rstruct}[2]{#1~\lstmath{\{}#2~\lstmath{\}}}
\newcommand{\Rblock}[1]{\lstmath{\{}\,#1\,\lstmath{\}}}
\newcommand{\Rblocke}[2]{\lstmath{\{} #1 \lstmath{;} #2 \,\lstmath{\}}}
\newcommand{\Rloop}[1]{\lstmath{loop}~#1}
\newcommand{\Rbreak}{\lstmath{break}}
\newcommand{\Rcontinue}{\lstmath{continue}}
\newcommand{\Rif}{\lstmath{if}}
\newcommand{\Relse}{\lstmath{else}}
\newcommand{\Rmatch}[2]{\lstmath{match}~#1~\lstmath{\{}#2~\lstmath{\}}}
\newcommand{\Rfncall}[2]{#1\lstmath{(}#2\lstmath{)}}
\newcommand{\Rfn}[1]{\lstmath{#1}}
\newcommand{\Rvar}[1]{\lstmath{#1}}
\newcommand{\Rlit}[1]{\lstmath{#1}}
\newcommand{\Rreturn}[1]{\lstmath{return}~#1}

\newcommand{\Rtrue}{\lstmath{true}}

\newcommand{\Rarm}[2]{#1~\lstmath{=>}~#2}
\newcommand{\Rarms}[2]{#1~\lstmath{,}~#2}

\newcommand{\WPx}[1]{\ensuremath{\text{wp}_\text{CF}~#1}}
\renewcommand{\iProp}{\text{iProp}}

\newcommand{\WPpat}{\ensuremath{\text{wp}_\text{pat}}}
\newcommand{\WPpats}{\ensuremath{\text{wp}_\text{pats}}}
\newcommand{\WPstmt}{\ensuremath{\text{wp}_\text{stmt}}}
\newcommand{\WPvexpr}{\ensuremath{\text{wp}_\text{vexpr}}}
\newcommand{\WPcexpr}{\ensuremath{\text{wp}_\text{cexpr}}}
\newcommand{\WPpexpr}{\ensuremath{\text{wp}_\text{pexpr}}}
\newcommand{\WPmove}{\ensuremath{\text{wp}_\text{move}}}
\newcommand{\WPtmp}{\ensuremath{\text{wp}_\text{tmp}}}
\newcommand{\WPuneg}{\ensuremath{\text{wp}_\text{uneg}}}
\newcommand{\WPbinop}{\ensuremath{\text{wp}_\text{binop}}}
\newcommand{\WPvexprs}{\ensuremath{\text{wp}_\text{vexprs}}}
\newcommand{\WPvalOver}{\ensuremath{\text{wp}_\text{val}}}

\newcommand{\WPblock}{\ensuremath{\text{wp}_\text{block}}}

\newcommand{\WPblockSealed}{\ensuremath{\text{wp}_\text{block}^\text{seal}}}
\newcommand{\WParm}{\ensuremath{\text{wp}_\text{arm}}}
\newcommand{\WParms}{\ensuremath{\text{wp}_\text{arms}}}
\newcommand{\WPfn}{\ensuremath{\text{wp}_\text{fn}}}

\newcommand{\delayedCpat}{\ensuremath{\text{delayed}_\text{cpat}}}

\newcommand{\framelet}[2]{#1@#2}

\newcommand{\FnItemOk}{\ensuremath{\text{fn\_item\_ok}}}

\newcommand{\bindCF}[0]{\ensuremath{\text{bind}_{\text{wp}_\text{CF}}}}
\newcommand{\retCF}[1]{\ensuremath{\text{ret}_{\text{wp}_\text{CF}}~#1}}
\newcommand{\nomatchCF}[1]{\ensuremath{\text{nomatch}_\text{CF}~#1}}

\DeclareMathOperator*{\mwand}{\mathrel{\text{-}\text{-}\text{-}\!\!\ast}}
\newcommand{\loc}{\ell}
\newcommand{\loctmp}{\ell_\text{tmp}}

\newcommand{\Rhrefdoc}[2]{\href{https://doc.rust-lang.org/reference/#1}{#2}\,\includegraphics[height=0.8em]{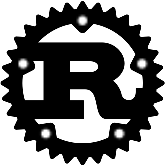}}

\newif\ifRhrefToDocs
\newcommand{\Rhref}[2]{\ifRhrefToDocs \hypertarget{ir:#2}{\href{https://doc.rust-lang.org/reference/#1}{#2}{\,\includegraphics[height=0.8em]{tex/figures/rust-logo.eps}}}
  \else \hyperlink{ir:#2}{\textsc{#2}}\fi
}

\newcommand{\WPVexprLiteralBool}{\Rhref{expressions/literal-expr.html\#r-expr.literal.bool}{{wp-vexpr-literal-bool}}}
\newcommand{\WPVexprLiteralInt}{\Rhref{expressions/literal-expr.html\#r-expr.literal.int}{{wp-vexpr-literal-int}}}
\newcommand{\WPStmtLet}{\Rhref{statements.html\#r-statement.let}{{wp-stmt-let}}}
\newcommand{\WPStmtExpr}{\Rhref{statements.html\#r-statement.expr}{{wp-stmt-expr}}}
\newcommand{\WPVexprBorrowValue}{\Rhref{expressions/operator-expr.html\#r-expr.operator.borrow.result}{{wp-vexpr-borrow-value}}}
\newcommand{\WPVexprBorrowPlace}{\Rhref{expressions/operator-expr.html\#r-expr.operator.borrow.temporary}{{wp-vexpr-borrow-place}}}
\newcommand{\WPPexprBorrow}{\Rhref{expressions.html\#r-expr.temporary}{{wp-vexpr-borrow-place}}}
\newcommand{\WPVexprUnOpNeg}{\Rhref{expressions/operator-expr.html\#r-expr.negate.results}{{wp-vexpr-unop-neg}}}
\newcommand{\WPVexprBinOp}{\Rhref{expressions/operator-expr.html\#r-expr.arith-logic.behavior}{{wp-vexpr-binop}}}

\newcommand{\WPPexprVar}{\Rhref{expressions/path-expr.html\#r-expr.path.intro}{{wp-pexpr-var}}}
\newcommand{\WPVexprVar}{\Rhref{expressions/path-expr.html\#r-expr.path.place}{{wp-vexpr-var}}}
\newcommand{\WPVexprTuple}{\Rhref{tuple-expr.html\#r-expr.tuple.value}{{wp-vexpr-tuple}}}
\newcommand{\WPPexprTupleField}{\Rhref{tuple-expr.html\#r-expr.tuple-index}{{wp-pexpr-tuple-index}}}
\newcommand{\WPVexprAdt}{\Rhref{expressions/struct-expr.html\#r-expr.struct.intro}{{wp-vexpr-adt}}}
\newcommand{\WPPexprAdtField}{\Rhref{expressions/struct-expr.html\#r-expr.struct.field}{{wp-pexpr-adt-field}}}
\newcommand{\WPVexprBlock}{\Rhref{expressions/block-expr.html\#r-expr.block.value}{{wp-vexpr-block}}}
\newcommand{\WPVexprIf}{\Rhref{expressions/if-expr.html\#r-expr.if.intro}{{wp-vexpr-if}}}
\newcommand{\WPVexprMatchV}{\Rhref{expressions/match-expr.html\#r-expr.match.scrutinee-value}{{wp-vexpr-match-v}}}
\newcommand{\WPVexprMatchP}{\Rhref{expressions/match-expr.html\#r-expr.match.scrutinee-place}{{wp-vexpr-match-p}}}
\newcommand{\WPVexprLoop}{\Rhref{expressions/loop-expr.html\#r-expr.loop.infinite}{{wp-vexpr-loop}}}
\newcommand{\WPVexprBreak}{\Rhref{expressions/loop-expr.html\#r-expr.loop.break}{{wp-vexpr-break}}}
\newcommand{\WPVexprContinue}{\Rhref{expressions/loop-expr.html\#continue-expressions}{{wp-vexpr-continue}}}
\newcommand{\WPVexprReturn}{\Rhref{expressions/return-expr.html\#r-expr.return.behavior}{{wp-vexpr-return}}}
\newcommand{\WPVexprCallGlobname}{\Rhref{expressions/call-expr.html\#r-expr.call}{{wp-vexpr-call-globname}}}

\newcommand{\WPBlockCons}{\textsc{wp-block-cons}}
\newcommand{\WPBlockNilExpr}{\textsc{wp-block-nil-e}}
\newcommand{\WPBlockNilNoExpr}{\textsc{wp-block-nil-ne}}
\newcommand{\WPPatConst}{\textsc{wp-pat-const}}
\newcommand{\WPIfJoin}{\textsc{wp-if-join}}

\DeclareMathOperator*{\pointsto}{\mapsto}
\DeclareMathOperator*{\pointstoq}{\mapsto_q}

\DeclareMathOperator*{\layoutq}{\texttt{layout}_q}
\DeclareMathOperator*{\bigsep}{\mbox{\LARGE$\ast$}}

\newcommand{\pointstoUninit}{\mapsto^\text{uninit}}
\newcommand{\pointstoUninitq}{\mapsto^\text{uninit}_q}

\DeclareMathOperator{\zip}{zip}
\DeclareMathOperator{\noover}{nooverflow}
\DeclareMathOperator{\cons}{:\!:}
\DeclareMathOperator{\nil}{[\;]}

\DeclareMathOperator{\guard}{\text{guard}}
\DeclareMathOperator{\throw}{\text{throw}}

\newcommand{\Seq}[1]{\text{seq}_{\text{CF}}\,#1}
\DeclareMathOperator{\Continue}{\text{continue}_\text{CF}}
\DeclareMathOperator{\Break}{\text{break}_\text{CF}}

\DeclareMathOperator{\PatNoMatch}{\text{nomatch}_\text{CF}}
\DeclareMathOperator{\MatchBreak}{\text{match}_\text{CF}}
\DeclareMathOperator{\Return}{\text{return}_\text{CF}}

\newcommand{\CFmonad}[1]{\ensuremath{\mathrm{CF}\,#1}}
\newcommand{\CFpost}[2]{\ensuremath{\{#1 . #2 \}}}

\newcommand{\ScopeNamed}{\ensuremath{\mathit{named}}}
\newcommand{\ScopeTmp}{\ensuremath{\mathit{tmp}}}

\newcommand{\WPdrop}{\ensuremath{\text{wp}_\text{drop}}}
\newcommand{\WPcallDrop}{\ensuremath{\text{wp}_\text{callDrop}}}
\newcommand{\WPguard}{\ensuremath{\text{wp}_\text{guard}}}
\DeclareMathOperator{\StorageLive}{\mathtt{StorageLive}}
\DeclareMathOperator{\StorageDead}{\mathtt{StorageDead}}
\DeclareMathOperator{\DropEvent}{\mathtt{Drop}}
\newcommand{\BreakN}[1]{\ensuremath{\text{break}_{\text{CF}}^{#1}}}
\newcommand{\ContinueN}[1]{\ensuremath{\text{continue}_{\text{CF}}^{#1}}}

\newcommand{\WPDropInit}{\textsc{wp-drop-init}}
\newcommand{\WPDropUninit}{\textsc{wp-drop-uninit}}

\newcommand{\seqK}{\ensuremath{\mathit{seqK}}}
\newcommand{\rtK}{\ensuremath{\mathit{rtK}}}
\newcommand{\scont}{\ensuremath{\mathit{scont}}}
\newcommand{\hcont}{\ensuremath{\mathit{hcont}}}
\newcommand{\SKnil}{\ensuremath{\mathit{SKnil}}}
\newcommand{\SKbind}{\ensuremath{\mathit{SKbind}}}
\newcommand{\HKnil}{\ensuremath{\mathit{HKnil}}}

\def\figuresize{\small}
\def\examplesize{\small}
 
\setboolean{showcomments}{false}

\begin{document}

\title{\sys{} - Foundational Veriﬁcation of Rust Programs}

\author{Franti\v{s}ek Farka}
\email{frantisek.farka@barkhauseninstitut.org}
\affiliation{
  \institution{ Barkhausen Institute}
  \country{Germany}
}

\author{Carmine Abate}
\email{carmine.abate@barkhauseninstitut.org}
\affiliation{
  \institution{Barkhausen Institute}
  \country{Germany}
}

\author{Sven Linker}
\email{sven.linker@kernkonzept.com}
\affiliation{
  \institution{Kernkonzept GmbH}
  \country{Germany}
}

\author{Sebastian Ertel}
\email{carmine.abate@barkhauseninstitut.org}
\affiliation{
  \institution{Barkhausen Institute}
  \country{Germany}
}
\email{sebastian.ertel@barkhauseninstitut.org}

\begin{abstract}

We present \textbf{\sys{}}, a foundational verification framework for Rust programs
  in the Rocq theorem prover, built on the Iris separation logic framework.
\sys{} provides the first semantics of surface-level Rust mechanized in a
  proof assistant with an attached program logic, directly grounded in
  the Rust Reference: it deeply embeds the Typed High-level Intermediate
  Representation (THIR) into Rocq and formalises Rust's dynamic
  semantics as a weakest-precondition predicate transformer calculus.
By operating at THIR rather than on internal compiler representations,
  \sys{} proof goals display the THIR AST, which pretty-prints to surface Rust,
  keeping verification close to the source code and facilitating maintainability
  as code evolves.
Atop this semantics, \sys{} develops a program logic and a
  syntax-directed proof automation layer; the program logic includes
  defunctionalized continuation stacks that keep proof goals first-order and compact.
Soundness is established incrementally, construct by construct, against an
  interaction-trees denotation.
A synthetic test suite demonstrates a two-to-four times reduction
  in proof size compared to raw semantic proofs.
  We further showcase \sys{} on a buddy allocator case study,
  verifying memory safety of the allocation and deallocation functions,
  laying the groundwork for end-to-end verification in a shared Rocq
  semantic foundation spanning hardware-software boundaries.
  
\end{abstract}



\maketitle

\begin{acks}
This work has been supported by funding from the Agentur für Innovation in der Cy-bersicherheit GmbH (Cyberagentur).
\end{acks}

\section{Introduction}

The Rust programming language has gained widespread adoption
in systems software~\cite{linux,SchuermannCGLLP25}, driven by its
combination of high performance and strong memory safety guarantees. At the heart
of the language lies a linear type system that enforces
ownership and borrowing rules~\cite{JungKD18},
preventing common memory safety errors such as use-after-free, dangling pointers,
and data races~\cite{MSRC2019Proactive,google2022memorysafe}.
\looseness=-1
However, memory safety alone is not sufficient for critical systems:
ensuring correct functional behavior, compliance with specifications, and
correctness of interactions between components requires formal verification that
accounts for both the behavior of individual components and their
interactions across system boundaries.
A framework that focuses in particular on end-to-end verification
of such critical systems is still missing.

\paragraph{The design space of existing Rust verification.}

Throughout the last decade, various verification frameworks for Rust programs
were proposed; each focusing on a different aspect in the design space.
Dynamic and bounded tools such as Miri~\cite{miri} and
Kani~\cite{kani} focus on detecting bugs, via concrete execution and bounded model
checking respectively.
SMT-based verifiers such as Verus~\cite{verus}, Prusti~\cite{prusti}, and
Creusot~\cite{creusot} focus on proof automation via external solvers.
Foundational approaches provide machine-checked soundness
proofs based on formal operational semantics,
supporting higher-order reasoning.
Aeneas~\cite{aeneas} shows that for a restricted fragment of the Rust language,
verification can be reduced to reasoning about a pure functional translation.
RefinedRust~\cite{refinedrust} combines a foundational Iris-based approach in Rocq with a high degree of proof automation.
Most of these approaches build on MIR, the mid-level
intermediate representation, which exposes borrowing information but discards source-program
structure, breaking the correspondence between proofs and code.

\paragraph{End-to-end verification for Rust-based systems.}
In a realistic systems stack a verified Rust component interacts with other components---software modules,
operating system services for example for networked I/O, or hardware
mechanisms such as CHERI capabilities~\cite{7163016,10.1145/3719027.3744861}.
For verification to scale,
proofs about software components need to stay close to the source code,
keeping verification maintainable as the code evolves~\cite{FMbedrock}, and
reasoning needs to be modular~\cite{GuSCWKSC16,KleinEHACDEEKNSTW09,Chlipala11}: the correctness of a caller
must follow from the specifications of its callees, without re-examining their implementations.
However, modular reasoning among Rust components alone does not extend across
system boundaries such as hardware.
Composing proofs across such boundaries requires a shared semantic foundation
in which the semantics of each layer can be related.

We present \textbf{\sys{}}, a foundational verification framework for
Rust programs in the Rocq theorem prover~\cite{rocq} that
addresses this gap.
\sys{} targets THIR, the Typed High-Level Intermediate Representation,
the first typed IR in the Rust compiler, which closely
follows the Rust language reference~\cite{rustref}.
\sys{} is built on Iris~\cite{iris}, a higher-order separation logic
framework, and deeply embeds Rust into Rocq via Hax~\cite{hax}.
This provides foundational guarantees, modular reasoning and a
semantic foundation for composition with other Rocq-verified components.
Unlike toy calculi, which typically omit control flow altogether, \sys{}'s
semantics accounts for Rust's expression-oriented control flow at the scale
of a full systems language.
By grounding the formal semantics in the Rust language reference,
we take a step toward opening the
definition of Rust's semantics to the broader programming language
research community.

\begin{figure}
  \figuresize
\definecolor{rustorange}{RGB}{211,69,22}
\definecolor{coqblue}{RGB}{113, 175, 243}
\begin{tikzpicture}[
    scale=0.15,
    font=\sffamily,
    node distance = 0.2cm,
    def/.style = {draw,rounded corners,align=center},
    box/.style={rectangle, minimum width=1.8cm, minimum height=0.5cm,
                  align=center, rounded corners=4pt},
    rust/.style={box, fill=rustorange!25, draw=white},
    layer/.style={box, minimum width=3.7cm, fill=coqblue, draw=white},
    intermediate/.style={box, fill=gray!15, draw=white},
    coqfile/.style={box, fill=coqblue!50, draw=white},
    itrees/.style={box, fill=coqblue},
    someitrees/.style={box, dotted, fill=coqblue!50, draw=white},
    noitrees/.style={box, draw, dashed, fill=white, thick},
]

\node[rust] (crate) at (0,0) {source crate};
\node[coqfile] (spec)     [right =1.5cm of crate] {specification};
\node[coqfile] (proof)    [right =0.5cm of spec] {proof};
\node[layer] (auto)     [right =1.25cm of proof] {automation (\S\ref{sec:auto})};
\node[layer] (proglog)  [below left=0.25cm and 0.75cm of auto.south east] {program logic (\S\ref{sec:proglog})};
\node[layer] (axsem)    [below left=0.25cm and 0.75cm of proglog.south east] {axiomatic semantics (\S\ref{sec:sem})};
\node[intermediate] (thirrocq) at (spec |- axsem) {THIR};
\node[intermediate] (thirrust) at (crate |- thirrocq) {THIR};

\node[itrees] (itreesem) [right =0.6cm of axsem] {\itree{}s (\S\ref{sec:itreemodel})};

\draw[-latex,thick] (crate.south) -- (thirrust.north)
node (hax) [midway, left] {hax};
\draw[-latex,thick] (thirrust.east) -- (thirrocq.west)
node (hax2rocq) [midway,above] {hax2rocq};
\draw[-latex,thick] (thirrocq.north) -- (spec.south);
\draw[-latex,dotted] (proof.west) -- (spec.east);
\path[name path=axsem--auto] (auto.south) -- (axsem.north); 
\path[name path=plnedge] (proglog.north east) -- (proglog.north west); 
\path[name path=plsedge] (proglog.south west) -- (proglog.south east); 

\path [name intersections={of=axsem--auto
            and plnedge,by=E}];
\path [name intersections={of=axsem--auto
            and plsedge,by=F}];
\draw[-latex,thick] (axsem.north) -- (F);
\draw[-latex,thick] (E) -- (auto.south);
\draw[-latex,thick] (auto.west) -- (proof.east)
node (guide) [midway,below] {\textit{guides}};

\draw[-latex,dotted] (axsem.west) -- (thirrocq.east)
node (denotethir) [midway,above] {\textit{denotes}};
\draw[-latex,dotted] (itreesem.west) -- (axsem.east);

\draw[draw=lightGray,thick] (hax2rocq.north) -- ([yshift=10cm]hax2rocq.north)
node (rust) [midway,left,yshift=-0.4cm] {\includegraphics[width=1.3em]{tex/figures/rust-logo.eps}}
node (rocq) [midway,right,yshift=-0.4cm] {\includegraphics[width=1.3em]{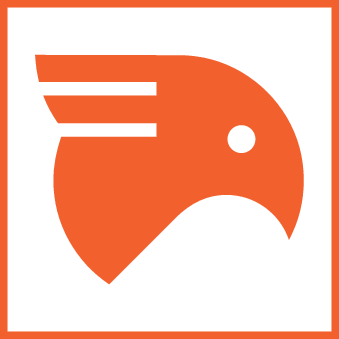}};

\draw[draw=lightGray,thick] ([yshift=-0.5cm]hax2rocq.south) -- ([yshift=-3cm]hax2rocq.south);

\end{tikzpicture}
\ff{``denotes'' label: I don't think it's a technically right term to use and we don't use it for axsem, but it was used for ITree denotations (which is gone/in Apx now)}
\ff{I am not sure what dashed ITree was intended to mean. My guess is ``in progress''?
I drew the half-filled node, but it seems like unfinished thing asking for trouble, any ideas how to change?}
   \caption{A \textcolor{rustorange}{Rust crate} is embedded via \texttt{hax} and
    \texttt{hax2rocq} as THIR in Rocq, on which the user writes a
    \textcolor{coqblue!50}{specification} and \textcolor{coqblue!50}{proof}, guided by
    the \sys{} framework's \textcolor{coqblue}{automation}, \textcolor{coqblue}{program
    logic}, and \textcolor{coqblue}{axiomatic semantics}, whose soundness is
    established via \textcolor{coqblue}{\itree{}s}.}
  \label{fig:arch}
\end{figure}

\paragraph{Contributions.}
The key contributions of our work are in both design and
implementation of \sys{}.
In this work we:

\begin{itemize}[leftmargin=*]
  \item provide an extensible architecture for foundational Rust
    verification: a layered design in which axiomatic semantics,
    program logic, and proof automation are cleanly separated, language
    coverage can be grown incrementally, and soundness can be
    established gradually as the framework matures (see Fig.~\ref{fig:arch}).
\end{itemize}
    This architecture is designed with industrial-scale verification
    as a goal, and we demonstrate it on a realistic case study as a
    first step toward it.
As concrete technical contributions, we:
\begin{itemize}[leftmargin=*]
\item propose an axiomatic weakest-precondition semantics for THIR,
  closely following the Rust Reference\footnote{Definitions marked with
  \includegraphics[height=0.8em]{tex/figures/rust-logo.eps}
  are linked to the Rust Reference in the electronic version of the paper.} and validate it against the intended semantics via a synthetic test suite;\item demonstrate that existing techniques~\cite{plalacarte} provide an incremental approach for
  instantiating the axiomatic semantics, and
  show how operational semantics and soundness proofs can be grown compositionally as language coverage expands;
\item implement a program logic layer exemplified by definition sealing, evaluation reordering,
  join-point reasoning, and defunctionalized continuation stacks;
\item build a syntax-directed proof automation layer atop the program logic and
        evaluate it on a synthetic test suite covering the language constructs supported by \sys{};
\item verify a buddy memory allocator~\cite{buddyalloc} comprising approximately 300 lines of Rust as a realistic case study.
\end{itemize}
The accompanying Rocq development supports a wider range of language features
than those presented in the paper, including trait method calls with
static dispatch and simple generics, further demonstrating the extensibility of the architecture
(remaining trait limitations are discussed in \S\ref{sec:relatedW}).
The full drop-scope calculus is out of scope for the main text and is given
in Appendix~\ref{apx:dropscopes}, which also discusses verification of
destructors for types with non-trivial \lstmath{Drop} implementations as
future work.

\paragraph{Structure of the paper}
In Sec.~\ref{sec:motivation}, we showcase \sys{} on a simple example
that illustrates both the expressiveness of Rust and the layers of
\sys{}'s framework; the example guides the reader through the
technical development and forward-references the formal definitions
introduced in later sections.
Sec.~\ref{sec:memorymodel} describes our memory model and the interface used to
capture value ownership.
In Sec.~\ref{sec:sem}, we present the axiomatic weakest-precondition
semantics and Sec.~\ref{sec:itreemodel} describes our approach to proving
their soundness with respect to an \itrees denotation in Sec.~\ref{sec:itreemodel}.
Sections~\ref{sec:proglog} and ~\ref{sec:auto} present
the program logic layer and the proof automation respectively.
In Sec.~\ref{sec:casestudy}, we report on the buddy allocator case study,
discuss related work in Sec.~\ref{sec:relatedW} and conclude in Sec.~\ref{sec:conclusion}.
 \section{\sys{} by Example}
\label{sec:motivation}

\noindent
What distinguishes Rust from other systems programming languages is that
Rust is primarily an \emph{expression} language,
thus providing programmers with a considerable degree of expressiveness.
Constructs such as control-flow expressions appearing as function call arguments
are idiomatic in Rust---for instance, an early return nested within the desugaring
of a conditional expression.
To illustrate, consider a simple, control-flow sensitive function \Rfn{example}
(Fig.~\ref{fig:motivation-function}), annotated with its functional specification.
\looseness=-1
The function takes a boolean argument and either produces the value 1 -- passing
it through a call to \Rfn{identity} -- or early returns 0,
with the branching expressed directly as a control-flow expression in the argument
position.

\sys{}'s automation yields a concise proof of the specification, the
$\mathit{example}_\mathit{ok}$ lemma (Fig.~\ref{fig:motivation-proof}).
The statement of the lemma reflects \sys{}'s modular approach to verification:
$\mathit{identity}_\mathit{spec},$ the specification of the callee \Rfn{identity}, entails the specification
of the caller, $ \mathit{example}_\mathit{spec}$ consisting of lines 1--4 of
Fig.~\ref{fig:motivation-function},
where the entailment ($\vdash$) is that of Iris separation logic\footnote{In the actual development, we state specifications separately from the code,
though we use Rocq notations to achieve the presentation shown above.}.
The tactics-based proof script mirrors the structure of the function itself (cf. Sec.~\ref{sec:auto}): to start the
verification of the function, the tactic \texttt{wp\_fn} handles the function prologue,
\texttt{wp\_call} reasons about the call to \Rfn{identity},
and $\texttt{wp\_if}~b$ reasons about the conditional
expression branching on $b$.
This syntactic correspondence between code and proof script is not a limitation of the framework but a conscious design choice that eases readability and maintainability of proofs.
In what follows, we discuss each proof step in detail, exposing the underlying Rust semantics and the layers of \sys{}'s framework that together enable this level of automation.
This necessarily forward-references formal definitions from later sections.\footnote{In
a modern PDF viewer, hovering over a cross-reference previews the referenced definition.}

\begin{figure}
  \figuresize
\centering
\begin{subfigure}{0.49\textwidth}
  \begin{lstlisting}
#[with(b)]
#[arg(bool b)]
#[post( v | if b then v = 1u8
              else v = 0u8 )]
pub fn example (b : bool) -> u8 {
  return (identity (
            if b { 1 }
            else { return 0; } )) }
\end{lstlisting}
    \caption{Rust program with specification annotations}
    \label{fig:motivation-function}
\end{subfigure}
\hfill
\begin{subfigure}{0.49\textwidth}
\begin{align*}
  \figuresize
  & \texttt{Lemma } \mathit{example}_\mathit{ok} :
\mathit{identity}_\mathit{spec} \vdash \mathit{example}_\mathit{spec}. \\
  & \texttt{Proof with } \mathtt{wp\_engage}. \\
  & \quad\mathtt{wp\_fn}\dots \\
  & \quad\mathtt{wp\_call}\dots \quad\text{(* call to \Rfn{identity} *)} \\
  & \quad\mathtt{wp\_if} ~ b\dots \\
  & \texttt{Qed}.
\end{align*}
    \caption{Automated Rocq proof (verbatim)}
    \label{fig:motivation-proof}
\end{subfigure}
\caption{Control-flow sensitive Rust function and its \sys{} proof}
\label{fig:motivation}
\end{figure}

Before starting the proof, \texttt{hax2rocq}, our extension to \textsc{Hax}~\cite{hax},
transpiles the Rust crate into a THIR AST in Rocq (Fig.~\ref{fig:arch}); all subsequent
reasoning operates on this AST.
\sys{}'s notion of function specification correctness (Def.~\ref{def:fn-item-ok}) relates the specification to the weakest precondition (WP) of the function definition
(\WPfn, Def.~\ref{def:wp-fn}),
located in the crate AST. The first proof step, tactic \texttt{wp\_fn}, performs the necessary lookups in the AST and reduces the $\WPfn$ goal to a WP of the function body. Interactively inspecting the proof state at this point, rather than relying on automation, reveals the following goal:

\newbox\bVarB
\setbox\bVarB\hbox{\Rvar{b}}
\newbox\bExBody
\setbox\bExBody\hbox{\Rreturn{\Rblock{\Rfncall{\Rfn{identity}}{\Rif~{\Rvar{b}}~{\Rblock{\Rlit{1}}~\Relse~{\Rblock{\Rreturn{\Rlit{0}}\lstmath{;}}}}}}}}

\begin{mathpar}
  \examplesize
  \mprset{flushleft}
  \inferrule[]
            {\mathit{identity}_\mathit{spec}
              \\
              \Phi[b] }
            {\xi \leftarrow \WPpat~\usebox\bVarB~(\valBool{b}) \nil~;
              \\\\
              \WPvexpr~\usebox\bExBody~\xi~;
              \\\\
              \guard \Return}

\end{mathpar}

The goal is an Iris entailment. The antecedent, the separation logic
(spatial) assumptions, comprise correctness of the specification of
the callee, $\mathit{identity}_\mathit{spec}$, and the postcondition
of the caller, abbreviated with $\Phi$---which is exactly the \texttt{post}
annotation on line 3 of Fig.~\ref{fig:motivation-function}:
\[
\Phi[b] = \lambda Q. \forall v. (v = \text{if } b \text{ then } 1 \text{ else } 0) \mwand Q~v
\]
The conclusion of the Iris entailment is a sequence of three weakest preconditions,
separated by the monadic bind ($\text{-} \leftarrow \text{-};$), in our control-flow-sensitive
specification monad (Sec.~\ref{sec:cfmon}).
In particular, the goal is accumulated from the WP of the \emph{identifier pattern}
in the function argument (\WPpat, Fig.~\ref{fig:wp-pat})
that introduces a new binding of local variable $\Rvar{b}$ into the function frame,
the WP of the function body, and the final guard asserting that function evaluation results in a return.
We \textcolor{syntaxblue}{highlight} the Rust AST for readability---
$\Rvar{b}$ is a Rust variable and $\valBool{b}$ encodes a boolean value
in our semantics, parameterized over the Rocq variable $b$.

Applying an inference rule of the predicate transformer calculus (Sec.~\ref{sec:sem})
to the conclusion of the entailment \emph{accumulates}
new proof obligations atop an implicit postcondition,
effectively reducing the goal step by step.
The inference rules are an underapproximation of the reference semantics,
ensuring that discharging all accumulated obligations is sufficient to establish
the original specification.

The semantics materialises local variable bindings in memory:
$\WPpat$ results in
the frame $\xi$ bound to $\framelet{\Rvar{b}}{addr_b}$,
denoting that the local variable is materialised at location $addr_b$,
and the corresponding points-to fact $addr_b \pointsto \valBool{b}$
is accumulated as an assumption, denoting that the location of $\Rvar{b}$
stores its value bound to $b$ in the ambient metalogic of Rocq.
The WPs of the return expression and the function call are then accumulated in turn.
The return expression (\WPVexprReturn{}, Fig.~\ref{fig:wp-vexpr2}) first accumulates
the WP of its operand and then performs the monadic $\Return$ operation~(Sec.~\ref{sec:cfmon}).
The call to $\Rfn{identity}$ (\WPVexprCallGlobname{}, Fig.~\ref{fig:wp-vexpr2})
decomposes the WP into the WP of the argument and
the $\WPfn$ of the function bound to the symbolic value produced by the argument:
\newbox\bExBodyA
\setbox\bExBodyA\hbox{\Rif~\Rvar{b}~{\Rblock{\Rlit{1}}~\Relse~{\Rblock{\Rreturn{\Rlit{0}}\lstmath{;}}{}}}}
\newbox\bExBodyAb
\setbox\bExBodyAb\hbox{\Rfn{identity}}
\newbox\bVarB
\setbox\bVarB\hbox{\Rvar{b}}
\begin{mathpar}
  \figuresize
    \mprset{flushleft}
    \inferrule[]
              {\mathit{identity}_\mathit{spec}
                \\
                \Phi[b]
                \\
                addr_b \pointsto \valBool{b} \
              }
              { v' \leftarrow \WPvexpr~\usebox\bExBodyA~[\framelet{\usebox\bVarB{}}{addr_b}]~;
                \\\\
                v \leftarrow \WPfn~\usebox\bExBodyAb~[v']~;
                \\\\
                \throw~(\Return~v) ~;
                \\\\
                \guard \Return
              }
\end{mathpar}

The $\WPfn$ of $\Rfn{identity}$ can be accumulated out of order;
the specification $\mathit{identity}_\mathit{spec}$ (cf. Def.~\ref{def:fn-item-ok})
reduces to $\retCF{v'}$ and can be canceled out
using the equational theory of the underlying monad.
Although the specification is persistent, we drop it from the context
as it is no longer needed.
The proof proceeds by case analysis on the value of $b$, yielding two subgoals.
The WP of each subgoal is accumulated according to the semantics of if
expressions (\WPVexprIf), simplifying each to the \WPvexpr of the corresponding branch:
\newbox\bExBodyB
\setbox\bExBodyB\hbox{\Rblock{\Rlit{1}}}
\newbox\bExBodyC
\setbox\bExBodyC\hbox{\Rblock{\Rreturn{\Rlit{0}}\lstmath{;}}{}}
\begin{mathpar}
    \figuresize
\mprset{flushleft}
    \inferrule[]
              {\Phi[true]
                \\
                addr_b \pointsto \valBool{true} \
              }
              { v \leftarrow \WPvexpr~\usebox\bExBodyB~[\framelet{\usebox\bVarB{}}{addr_b}]~;
                \\\\
                \throw~(\Return~v)~;
                \\\\
                \guard \Return
              }

    \inferrule[]
              {\Phi[false]
                \\
                addr_b \pointsto \valBool{false} \
              }
              { v \leftarrow \WPvexpr~\usebox\bExBodyC~[\framelet{\usebox\bVarB{}}{addr_b}]~;
                \\\\
                \throw~(\Return~v)~;
                \\\\
                \guard \Return
              }
\end{mathpar}
The left subgoal follows from the semantics of blocks (\WPblock, Def.~\ref{def:wp-block})
and integer literals (\WPVexprLiteralInt),
producing a symbolic value that is then bound in the $\Return$ operation.
The other subgoal is more interesting: the WP of the block is accumulated,
followed by the WP of the expression statement (\WPStmtExpr),
arriving at the return expression.
As described before, the WP of the operand--the literal $\Rlit{0}$ in this case--is first accumulated,
and then the $\Return$ operation is sequenced.
After simplification, we arrive at the following two goals:
\newbox\bExBodyD
\setbox\bExBodyD\hbox{\lstmath{u8}}
\begin{mathpar}
  \figuresize
\mprset{flushleft}
  \inferrule[]
            {\Phi[true]
              \\
              addr_b \pointsto \valBool{true} \
            }
            { \throw~(\Return~1)~;
              \\\\
              \guard \Return
            }

            \inferrule[]
                      {\Phi[false]
                        \\
                        addr_b \pointsto \valBool{false} \
                      }
                      { n <- \throw~(\Return~0)~;
                        \\\\
                        \throw~(\Return~n) ~;
                        \\\\
                        \guard \Return
                      }
\end{mathpar}
The $\guard$ in
the left subgoal simply traps the $\Return$ and reverts control flow
into normal state while binding its result to the operand of $\Return$,
and is closed by the assumption
$\Phi[true]$.
The right subgoal contains two $\Return$ operations:
the return expression is itself an expression, technically yielding a value of
type \emph{never} (\lstmath{!}).
However, the monadic bind skips over the remainder of the sequence after the first
$\Return$, and the $\guard$ receives the value 0 of the first return.
The goal is then closed using $\Phi[false]$, concluding the proof.

The running example demonstrates several points.
The semantics of Sec.~\ref{sec:sem} is given directly, staying close to the
Rust Reference, but is not always the most convenient basis for reasoning.
Large goals reduce theorem prover performance and lead to long Qed times.
The block definition \WPblock{} simplifies to the constituent statement WPs,
accumulated one closure per bind step by the $\bindCF$ combinator; for longer
blocks this remains manageable, as the closures stay sequential.
The real source of blowup is nested control flow: the case analysis on $b$ above
duplicates the continuation and the proof of the remainder of the function across
both branches, compounding with nesting depth and pushing Qed times to minutes.
The out-of-order proof of \WPfn{} for $\Rfn{identity}$ also requires extra steps.
To address these concerns, we build a layer of program logic atop the base semantics;
in Sec.~\ref{sec:proglog}, we concretely showcase definition sealing,
evaluation reordering, join-point reasoning,
and defunctionalized continuation stacks (Sec.~\ref{subsec:defuncont}),
each addressing one of the issues raised above, in order.

This pattern of control flow nested in argument position, exercised by our running
\Rfn{example}, arises naturally in real-world Rust.
Chaining method calls with the question-mark operator is a compelling instance:
the following line from our buddy-allocator case study
manipulates block indices using checked arithmetic:
\begin{lstlisting}
  blocks.checked_add(offset)?.checked_shl(block_order)
\end{lstlisting}
Though concise and verifiable by \sys{}, this idiomatic code is
surprisingly complex upon desugaring, nesting \Rfn{checked_add} in the
argument of \Rfn{checked_shl} and involving several definitions from \Rfn{core::ops}.

These syntax-directed proof steps admit simple yet powerful automation, which the
program logic layer preserves and we build upon (Sec.~\ref{sec:auto}).

 \section{Memory Model}\label{sec:memorymodel}

To reason formally about Rust programs, we need a memory model that captures how
values are manipulated in the Rust Abstract Machine, the idealized model underlying Rust's semantics;
our work is inspired by MiniRust~\cite{minirust}.
For the vast majority of Rust programs---even systems code---functional correctness depends on
\emph{which values} are stored and \emph{who owns them}, not on their byte-level
representation.
Verifying that a buddy allocator tracks free blocks, for example, requires
ownership of those blocks but not their byte encoding.
By contrast, unsafe code that peeks through value representations---via
\lstmath{transmute}, lock-free atomics, or memory-mapped I/O registers---does
depend on the concrete representation, but forms a small, well-identified subset.
We exploit this by modelling memory as an abstract interface
that captures ownership using fractional permissions and
organises it as a forest over basic values:
structured-type values are represented as trees,
with elements and fields accessible via index and field offsets and
their layout kept abstract.\footnote{We have proven this forest-of-values
interface atop an abstract byte-level memory model (single-threaded); its
description is out of scope for this paper.}

\subsection{Values and Locations}

\begin{figure}
\figuresize
  \begin{align*}
  \RTy \ni r ::=~&
  \RTyInt{s}{w} \mid
  \RTyBool \mid
  \RTyPtr \mid
  \RTyTuple{\overline{r}} \mid
  \RTyStruct{d} \mid
  \RTyEnum{d}
  \\
  \Loc \ni \loc ::=~& \locBase{\RTy} \mid
    \locField{\loc}{d} \mid
    \locIx{l}{n} \mid
    \locConst{d}
  \\
  \Val \ni v ::=~&
  \valInt{s}{w}{z} \mid
  \valBool{b} \mid
  \valPtr{\loc} \mid
  \valTuple{\overline{(n_i, v)}} \mid
  \valStruct{d}{\overline{(d_i, v)}} \mid
  \valEnum{d}{d'}{\overline{(d_i, v)}}
\end{align*}
  \caption{Values, runtime types, and locations of Rust Abstract Machine}
  \label{fig:valrtyloc}
\end{figure}

Values are either of basic types---integers, booleans, pointers, and references---or
structured types---tuples, structs, and enumerations---as captured by the \emph{runtime type} $\RTy$ (Fig.~\ref{fig:valrtyloc}).
We represent Rust fixed-size arrays as homogeneous tuples. We write $\overline{r}$ for a sequence of elements of $\RTy$---a notation we use
for sequences throughout the paper---so $\RTyTuple{\overline{r}}$ is a tuple type with field types drawn from $\overline{r}$.
Definition identifiers $d$ range over fully qualified names of source Rust elements;
in $\RTyStruct{d}$ and $\RTyEnum{d}$, $d$ identifies the type in a source crate.
We further use local identifiers $id$ for names of local variables.

Variables and intermediate results of evaluation may be dereferenced through their \emph{locations}
to access their values, or the elements of tuples and fields of structured values.
Locations $\Loc$ are
either roots $\locBase{\RTy}$---base allocations on the stack, heap, or global objects---or
reachable from a root via index ($\locIx{l}{n}$) or field ($\locField{l}{d}$) offsets;
they are thus paths through the memory forest.

Integer values $\valInt{s}{w}{z}$ are indexed with signedness
$s \in \{\mathit{usgn}, \mathit{sgn}\}$ and bit width
$w \in \{8, 16, 32, 64, 128, \lstmath{size}\}$,
with $z \in \mathbb{Z}$ drawn from the ambient logic.
Booleans $\valBool{b}$ are likewise modeled by ambient-logic booleans,
and pointers $\valPtr{\ell}$ by a location.
Tuples $\valTuple{\overline{(n, v)}}$ carry a finite map from indices
to values, encoded as a sequence of pairs; uniqueness of indices is
validated during parsing and maintained by the dynamic semantics of Sec.~\ref{sec:sem}.
Similarly, $\valStruct{d}{\overline{(d_i, v)}}$ and
$\valEnum{d}{d'}{\overline{(d_i, v)}}$ carry finite maps from fields
to values, indexed additionally by type-definition and constructor name.
There is an obvious forgetful mapping $\toRTy{\text{-}} : \Val \to \RTy$ from values to runtime types.

\subsection{Fractional Abstract Memory Model}

\begin{figure}
  \figuresize
  \begin{align*}
\loc \pointsto v ::=
    \loc \mapsto^{\valInt{s}{w}{}}_{q} z \mid
    \loc \mapsto^{\valBool{}}_{q} b \mid
    \loc \mapsto^{\valPtr{}}_{q} \loc' \mid
    \layoutq r \mid
    \loc \pointstoUninitq r
    \end{align*}
  \caption{\emph{points-to} relation for basic types}
  \label{fig:points-to}
\end{figure}

Fractional permissions~\cite{Boyland03,BornatCOP05} are a standard
separation-logic tool; we use fractional \emph{points-to}
assertions to reason precisely about shared and mutable accesses to
values in memory.  The interface ($\text{-}
\pointstoq \text{-}$) is shown in Fig.~\ref{fig:points-to}.  The relation
captures core data types, i.e.  integers ($\loc
\mapsto^{\valInt{s}{w}{}}_{q} z$), booleans ($\loc
\mapsto^{\valBool{}}_{q} b$), and pointers ($\loc
\mapsto^{\valPtr{}}_{q} \loc'$) as well as the memory \emph{layout} of
structures and enumerations; tuples need none, being laid out
sequentially in memory.  We also represent
uninitialized values ($\loc \pointstoUninitq r$), indexed by their
runtime type.  Fractions are drawn from $\{ q \in
\mathbb{Q} \mid 0 < q \leq 1 \}$, dropping $q$ when it equals~$1$.
\looseness=-1
Importantly, this interface abstracts over the concrete memory representations,
the layout abstracts the ownership of any padding bytes.
Using this interface, we define a points-to relation for values:

\begin{definition}[Points-to relation, $\text{-}\pointstoq\text{-}$]
{\figuresize
\begin{align*}
  \loc \pointstoq \valInt{s}{w}{z} \triangleq~& \loc \mapsto_{\valInt{s}{w}{}} z
&   \loc \pointstoq \valStruct{d}{fs} \triangleq~&
    \layoutq \toRTy{\valStruct{d}{fs}}
\\
  \loc \pointstoq \valBool{b} \triangleq~& \loc \mapsto_{\valBool{}} b
& & \quad \ast \bigsep_{ (f, v) \in fs} (\loc . f \pointstoq v ) \\
  \loc \pointstoq \valPtr{\loc'} \triangleq~& \loc \mapsto_{\valPtr{}} \loc'
& \loc \pointstoq \valEnum{d}{d'}{fs} \triangleq~&
    \layoutq \toRTy{\valEnum{d}{d'}{fs}}
  \\
  \loc \pointstoq \valTuple{vs} \triangleq~& \bigsep_{ (i, v) \in vs} ( \loc . i \pointstoq v )
  & & \quad \ast \bigsep_{ (f, v) \in fs} (\loc . f \pointstoq v )
\end{align*}
}
\end{definition}

\subsection{Borrowing and Unsafe Rust}

\looseness=-1
In Rust, borrows are the compiler's primary mechanism for enforcing the
linear type system~\cite{WagnerGML025} and safe memory access.
From a verification perspective, they are a sound but incomplete
approximation of ownership-safe memory manipulations: they guarantee that
all compiler-checked code respects the memory model, but not every valid
manipulation a programmer might perform---so unsafe blocks allow operations
outside the borrow checker's decidable fragment.
Our memory model handles both uniformly: safe code is code whose ownership
discipline the compiler discharges statically, whereas unsafe code discharges
it manually.
Within safe Rust, borrow-checked code can often be verified automatically,
reflecting the linear type discipline.

For unsafe code instead, the verification engineer interacts with the
theorem prover to manually split ownership as needed by complex memory
access patterns while maintaining soundness with respect to the underlying
memory model (see Sec.~\ref{sec:proglog} for the program logic layer that supports this).

A borrow's end is not modeled as a separate temporal event but falls out of
the spatial resource discipline: because the encoding logic is affine, a WP
rule that consumes an ownership fact via the separating conjunction and only
returns a (possibly different) fact via the magic wand encodes a temporal
transition as a spatial exchange---$\WPmove$ (Sec.~\ref{sec:movecopy}) is the
paradigmatic instance, exchanging $\loc \pointsto v$ for
$\loc \pointstoUninit \toRTy{v}$. This coincides with the discipline Tree
Borrows~\cite{VillaniHDJ25} articulates operationally: a borrow's lifetime
ends at the next conflicting access.\footnote{Appendix~\ref{apx:borrows} exposes
a mutable-reborrow example through the pattern rules.}
The axioms under-approximate the Reference semantics: programs violating the
ownership discipline yield unprovable goals, e.g.\ $\WPmove$ demands
$\loc \pointsto v$. For the adequacy-covered fragment (Sec.~\ref{sec:itreemodel})
this is a semantic guarantee; beyond it, it is enforced proof-theoretically.

The borrow checker is outside the TCB: the dynamic semantics consumes no
borrow-checker facts, so no proof's soundness depends on its
correctness---it is an untrusted oracle whose acceptance merely predicts
that the fractional obligations of safe code are dischargeable, as in practice
they often are. rustc's parsing, name resolution, and type checking remain in
the TCB. This contrasts with solver-based tools (Verus, Prusti, Creusot) that
assume well-borrowed input; foundational MIR-level tools such as
RefinedRust~\cite{refinedrust} likewise do not trust the borrow checker, but
encode the discipline explicitly in a lifetime logic. Our design instead
leaves it implicit: on borrow-checked code, the checker's untrusted acceptance
is precisely what makes our automation's fractional obligations dischargeable
without any lifetime reasoning of our own.
 \section{Axiomatic Semantics}
\label{sec:sem}

\begin{figure}
  \figuresize
  \begin{align*}
\Type \ni ty ::=~& \Rbool \mid \Rchar \mid \Rnum{\mathit{sgn}}{w} \mid 
    \lstmath{&} m ~ty \mid \Rnever \mid \\
    & \Rtuple{\overline{ty}} \mid
    \Rfun{\overline{ty}}{ty} \mid \Rarray{ty}{ce} \mid \Rslice{ty} \mid \Radt{d} \\
\Pat \ni pat ::=~& \Rpatwild \mid \lstmath{ref}^?~\lstmath{mut}^?~v \mid ce \mid
    \Rref\lstmath{mut}^?~pat \mid \Rnever \mid
\Rtuple{\overline{ty}} \mid
    \Rslice{\overline{ty}} \mid
    \Radt{d}~\overline{pat}
\\
    \Stmt \ni stmt ::=~& e \mid \lstmath{let}~pat~( = e )^?
    \\
    \Expr \ni e ::=~& lit \mid v \mid gn \mid \Rref \lstmath{mut}^? e \mid
      \Rderef e \mid \lstmath{-}e \mid \lstmath{!}e \mid
      e \odot e \mid
      \Rcast{e}{ty} \mid e~\lstmath{=}~e \mid
      \\ &
      \Rslice{\overline{e\lstmath{,}}}
      \mid \Rarray{e}{ce} \mid
\Rtuple{\overline{e\lstmath{,}}} \mid e.n \mid
      \Rstruct{gn}{\overline{gn~ (\lstmath{:} e)^?} } \mid e\lstmath{.}gn \mid
      \\
      & \Rblock{\overline{stmt\lstmath{;}}~e^?} \mid \Rloop{e} \mid \Rbreak \mid \Rcontinue \mid
      \\
      & \Rif~e~(\Relse~e)^? \mid
      \Rmatch{e}{\overline{pat~\lstmath{=>}~e\lstmath{,}}} \mid
      \\
      & \Rfncall{gn}{\overline{e\lstmath{,}}} \mid \Rreturn{e^?}
  \end{align*}
  \caption{Syntax of Rust: Types, patterns, statements, and expressions}
  \label{fig:syntax-ty}
\end{figure}

\sys{} leverages the frontend of the Hax tool, which interfaces directly with \texttt{rustc} via a compiler driver,
to obtain an AST representation of the program syntax in Rocq.
By relying on the compiler for, e.g., parsing, name resolution, and type checking, we delegate the static semantics to the compiler;
these components are part of the trusted computing base (TCB),
while all dynamic semantics reasoning and proofs of correctness are carried out within our verified framework and are thus outside the TCB. 
\looseness=-1 The syntax (Fig.~\ref{fig:syntax-ty}) in this paper is displayed in a style that resembles surface level
Rust; we give a full account in Appendix~\ref{sec:syn}.

We formalize \emph{dynamic semantics} of Rust in the form of inference
rules of a weakest-precondition predicate transformer calculus.
The calculus is structured in two layers:
a set of core transformers for value expressions (\WPvexpr),
place expressions (\WPpexpr), statements (\WPstmt), and
constant expressions (\WPcexpr) that are parameters of a Rocq module type
and instantiated on top of an \itrees model for soundness (Sec.~\ref{sec:itreemodel}), 
and a set of derived predicates, such as those for functions (\WPfn), patterns (\WPpat), and
blocks (\WPblock), expressed in terms of the core ones.
The axioms of the module type, covering both core and defined predicates,
closely follow the Rust Reference,
give the predicate transformations,
and act as a \emph{proof interface} rather than an executable model;
we validate them against the intended semantics via
a synthetic test suite (cf. Sec.~\ref{sec:auto}),
checking that each supported construct behaves as expected.

\subsection{Control Flow and Specification Monad}
\label{sec:cfmon}

Evaluation of an expression can have six possible outcomes:
a value is produced and evaluation continues sequentially;
a \lstmath{return} expression terminates the current function frame passing its operands value;
a \lstmath{break} expression terminates the corresponding loop body;
a \lstmath{continue} expression resumes evaluation at the beginning of the loop body;
a pattern match fails;
and a completed \lstmath{match} expression passes control to the remainder of the program.
These outcomes are uniformly described by the following type:

\begin{definition}[Control flow, $\CFmonad{\cdot}$]\label{def:CF}
\[
\CFmonad{A} :=
  \Seq{(a : A)} \mid \Return{v} \mid \Break \mid \Continue \mid \PatNoMatch \mid \MatchBreak
\]
\end{definition}
\noindent
$\Break$ targets the innermost enclosing loop and $\Continue$ resumes its
current iteration; \sys{} does not yet support labeled
\lstmath{break}/\lstmath{continue} (Sec.~\ref{sec:expr}), which would require
extending these constructors with a depth or label index.
Accumulating weakest precondition as shown in Sec~\ref{sec:motivation}
gives rise to a \emph{continuation} monad $\mathit{Cont}_R(A) \triangleq (A \to R) \to R$.
The type $\CFmonad{\cdot}: \mathrm{Type} \to \mathrm{Type}$ classifies
the state of the computation---$\Seq{A}$ as non-exceptional computations, and
the other constructors as the exceptional computation $E$
corresponding to the evaluation of control-flow transferring expressions.
This classification gives rise to an \emph{exception monad} $\CFmonad{A} \cong A + E$ and the respective transformer.
The bind and return of \CFmonad{\cdot} are those of the
exception monad, i.e. the bind concatenates sequential executions while
propagating exceptions, while the other constructors trigger an exception.

Applying the exception monad transformer to the continuation monad over Iris's $\iProp$
yields the \emph{control-flow sensitive specification monad} $\WPx{\cdot}$;

\begin{definition}[Control-flow sensitive specification monad, core transformers and guard]\label{def:corepreds}
\begin{align*}
    \WPx{A} & \triangleq (\CFmonad{A} -> iProp) -> iProp \\
\guard :~& \forall A\; (c : \CFmonad{A}).\;\WPx{A} \to (\CFmonad{A} -> \WPx{A}) -> \WPx{A} \\
\throw :~& \forall A\; (c : \CFmonad{A}).\;\WPx{A}
  \end{align*}
\begin{align*}
  \WPvexpr :~& \Expr \to \Xi \to \WPx{\Val}
& \WPcexpr :~& \Expr \to \WPx{\Val} \\
  \WPpexpr :~& \Expr \to \Xi \to \WPx{\Loc}
& \WPstmt :~& \Stmt \to \WPx{\Xi} \\
\end{align*}
\end{definition}
\noindent
The monad $\WPx{(\cdot)}$ enjoys the desirable properties of a
\emph{specification monad}~\cite{d4all, maillardthesis}.
After partial application, e.g. $\WPstmt (s : \Stmt)$, the core
weakest-precondition transformers in the definition above satisfy the
equational theory of the $\WPx{(\cdot)}$ monad.
As a consequence, all the derived weakest-precondition transformers
discussed in this paper satisfy the equational theory of specification
monads.
Finally, $\throw A\;c\;Q = Q(\mathit{inr}\;c)$ forces the evaluation
of a $Q : \CFmonad{A} -> iProp$ in the exceptional context $c$, 
and $\guard$ provides a handler to resume sequential
evaluation\footnote{we omit $A$ in the rest of the paper and just
write $\throw\;c, \guard\;c.$}.
\ff{todo: explain guard}

\subsection{Patterns}\label{sec:pat}

\begin{figure}
  \figuresize
  \begin{flushleft}
    \boxed{$$\WPpat : \Pat \to \mathcal{V} \to \Xi \to  \WPx{\Xi}$$}
  \end{flushleft}

  \begin{align*}
\WPpat ~ \Rpatwild ~ v ~ \xi &~\triangleq~ \retCF{\xi} \\
\WPpat ~ (\lstmath{mut}~i) ~ v ~ \xi &~\triangleq~
      \forall \loctmp, \loctmp \mapsto v \mwand \retCF{(\xi \cup \framelet{i}{\loctmp})}
      \\
      \WPpat ~ i ~ v ~ \xi &~\triangleq~
      \forall \loctmp ~ q, \loctmp \mapsto_q v \mwand \retCF{(\xi \cup \framelet{i}{\loctmp})}
      \\
      \WPpat ~ (\lstmath{ref}~\lstmath{mut}~i) ~ v ~ \xi &~\triangleq~
      \exists \loc, \loc \mapsto v \ast (\loc \mapsto v \mwand
      \\
      & \quad\quad\quad \forall \loctmp, \loctmp \mapsto \lstmath{ptr}~\loc \mwand \retCF{(\xi \cup \framelet{i}{\loctmp})}
      \\
      \WPpat ~ (\lstmath{ref}~i) ~ v ~ \xi &~\triangleq~
      \exists \loc~q, \loc \mapsto_q v \ast (\loc \mapsto_q v \mwand
      \\
      & \quad\quad\quad \forall \loctmp~q, \loctmp \mapsto \lstmath{ptr}~\loc \mwand_q \retCF{(\xi \cup \framelet{i}{\loctmp})}
      \\
\WPpat ~ (\Radt{d~pats}) ~ v ~ \xi &~\triangleq~
      \text{if}~ v = \Radt{d~pats} ~\text{then}~
\WPpats~pats~\xi \\
      & \quad\quad \text{else}~
\throw \nomatchCF
\\
\\
      \WPpats ~ (\Radt{d~pat \cons pats}) ~ v \cons vs ~ \xi &~\triangleq~
      \xi' \leftarrow \WPpat~pat ~ v ~ \xi  ~;
~ \WPpats~pats~vs~\xi' \\
      \WPpats ~ (\Radt{d~\nil}) ~ \nil ~ \xi &~\triangleq~ \retCF{\xi} \\
  \end{align*}
\caption{Weakest precondition predicate for patterns, \Rhrefdoc{patterns.html\#r-patterns}{\WPpat}}
  \label{fig:wp-pat}
\end{figure}

The weakest precondition semantics of patterns, $\WPpat$ is given in Fig.~\ref{fig:wp-pat};
we discuss destructuring patterns and identifier patterns in particular.

\paragraph{Destructuring patterns and refutability}
Patterns can be used to destructure values of struct, enum, and tuple type into their components.
Semantically, this amounts to checking that the type of the value corresponds to the type of the pattern
and delegating component values to sub-patterns, as shown for the $\Radt{d~pats}$ case in Fig.~\ref{fig:wp-pat}.
Note that, when the pattern does not match, the control flow transitions to $\PatNoMatch{}$.
That makes the pattern refutable, and refuting the pattern transitions the control flow.
This works also under a chain of sub-patterns as these are sequentialised after each other.
\ff{TODO: should we elaborate more on refutable patterns?}

\paragraph{Identifier patterns}
bind a value to a variable. By default, identifier patterns copy or move the value (\emph{cf.} Sec.~\ref{sec:movecopy}),
but bind by reference or mutable reference if annotated accordingly.
Our semantics distinguishes the individual cases, requires that the sufficient ownership of
points-to facts is available, and introduces the new points-to fact for the bound variable.

\subsection{Statements}\label{sec:stmt}

\begin{figure}\RhrefToDocstrue
  \figuresize

\begin{flushleft}
    \boxed{$$\WPstmt : \Stmt \to \Xi \to  \WPx{\Xi}$$}
  \end{flushleft}

  \newbox\bLet
  \setbox\bLet\hbox{\lstmath{let} pat \lstmath{=} e}
  \begin{mathpar}
    \mprset{flushleft}
    \inferrule[\WPStmtLet]
              {v \leftarrow \WPvexpr~e~\xi~;
                \\\\
                \WPpat~pat~v~\xi }
{\WPstmt~(\usebox\bLet{})~\xi}

    \inferrule[\WPStmtExpr]
              {\WPvexpr~e~\xi\qquad e \in\Expr }
              {\WPstmt~e~\xi}
  \end{mathpar}
  
  \crmn{please double-check my changes to \WPStmtExpr}
  
  \caption{Weakest precondition predicate for statements, $\WPstmt$}
  \label{fig:wp-stmt}
\end{figure}\RhrefToDocsfalse

Statements in Rust serve primarily to sequentialize expression evaluation.

\paragraph{Let statement}
Given a pattern, a let declaration introduces a set of variables.
The variables are initialized based on an optional initializer expression, or uninitialized otherwise.
For the sake of simplicity, we support only let statements with initializer expression.
The corresponding axiom is \WPStmtLet{} in Figure~\ref{fig:wp-stmt}.
The statements asserts that with function frame $\xi$, the weakest precondition of a let statement
with pattern $pat$ and initializer expression $e$ is entailed by sequentially obtaining the
weakest precondition of expression $e$, yielding a value $v$, and then obtaining
the weakest precondition of matching the pattern $pat$ to the yielded value.
Note that $\WPpat~pat~v$ is a weakest precondition over function frames $\WPx{\Xi}$.
That indeed captures the intuition, the let statement enriches function frames with
bindings captured by the pattern.

\paragraph{Expression statement}
Expression statement evaluates an expression and drops the result.
Thus, the weakest precondition of an expression statement \WPStmtExpr{} (Fig. \ref{fig:wp-stmt}) is the
weakest precondition of the expression in the current frame, $\WPvexpr~e~\xi$ followed by
simple return of the unchanged function frame, $\retCF~\xi$.

\subsection{Expressions}\label{sec:expr}

\begin{figure}\RhrefToDocstrue
  \figuresize
  \begin{flushleft}
    \boxed{$$\WPvexpr : \Expr \to \Xi \to  \WPx{\Val}$$}
  \end{flushleft}

  \newbox\bTrue
  \setbox\bTrue\hbox{\lstmath!True!}
  \newbox\bFalse
  \setbox\bFalse\hbox{\lstmath!False!}
  \newbox\bNum
  \setbox\bNum\hbox{\lstmath!num!}
  \newbox\bRef
  \setbox\bRef\hbox{\Rref{e}}
  \begin{mathpar}
    \mprset{flushleft}
    \inferrule[\WPVexprLiteralBool]
              {\retCF{(\valBool{lit})}  \and lit \in \{ \usebox\bTrue{}, \usebox\bFalse{} \} }
{\WPvexpr~lit~\xi}

    \inferrule[\WPVexprLiteralInt]
              {\retCF (\valInt{sng}{w}{lit}) \and lit : \usebox\bNum{}~sgn~w }
{\WPvexpr~lit~\xi}

    \inferrule[\WPVexprVar]
              {\ell \leftarrow \WPpexpr~id~\xi ~;
                \\\\
                \WPmove~\ell
              }
{\WPvexpr~id~\xi}

    \inferrule[\WPVexprBorrowPlace]
              {\ell \leftarrow \WPpexpr~e~\xi ~;
                \\\\
                \retCF{(\valPtr{\ell})} \and \quad e \text{ is pexpr}
              }
{\WPvexpr~\usebox\bRef{}~\xi}

    \inferrule[\WPVexprBorrowValue]
              {v \leftarrow \WPvexpr~e~\xi ~;
                \\\\
                \loc \leftarrow \WPtmp~v
                \\\\
                \retCF{\valPtr{\loc}} \and \quad\quad\qquad e \text{ is vexpr}
              }
{\WPvexpr~\usebox\bRef{}~\xi}

    \inferrule[\WPVexprUnOpNeg]
              {v \leftarrow \WPvexpr~e~\xi ~;
                \\\\
                \WPuneg~\neg~v
              }
{\WPvexpr~\neg e~\xi}

    \inferrule[\WPVexprBinOp]
              {v_l \leftarrow \WPvexpr~e_l~\xi ~;
                \\\\
                v_r \leftarrow \WPvexpr~e_r~\xi ~;
                \\\\
                \WPbinop~\odot~v_l~v_r
              }
{\WPvexpr~e_l \odot e_r~\xi}

  \end{mathpar}

  \caption{Weakest precondition predicate for value expressions, $\WPvexpr$}
  \label{fig:wp-vexpr}
\end{figure}\RhrefToDocsfalse

\begin{figure}\RhrefToDocstrue
  \figuresize
  \begin{flushleft}
    \boxed{$$\WPpexpr : \Expr \to \Xi \to  \WPx{\Loc}$$}
  \end{flushleft}

  \newbox\bRef
  \setbox\bRef\hbox{\Rref{m~e}}
  \begin{mathpar}
    \mprset{flushleft}
    \inferrule[\WPPexprVar]
              {\retCF{\loc}
                \and \xi[id] = \ell }
{\WPpexpr~id~\xi}

    \inferrule[\WPPexprBorrow]
              {v \leftarrow \WPvexpr~\usebox\bRef{}~\xi ~;
                \\\\
                \WPtmp~v
              }
{\WPpexpr~\usebox\bRef{}~\xi}

  \end{mathpar}

  \caption{Weakest precondition predicate for place expressions, $\WPpexpr$}
  \label{fig:wp-pexpr}
\end{figure}\RhrefToDocsfalse

The syntactic category $\Expr$ of expressions represents the core of the Rust language.
An expression always evaluates to a value, and it may have side effects.
Expressions are nested within each other, built up from sub-expressions, called \emph{operands}, and
their structure determines the control flow (cf. Sec.~\ref{sec:motivation}).

\paragraph{Place and value expressions}
Expressions fall into two categories: place expressions (pexpr)---paths referring
to variables, dereferences, indexing, and field expressions, and denoting a memory location---and
value expressions (vexpr), the rest, denoting actual values of the Rust Abstract Machine (\Cref{sec:memorymodel}).

Within each expression, sub-expressions may occur in \emph{place} or \emph{value expression context}.
For example, the operand of a field expression occurs in place expression context.
Our semantics captures the distinction between place and value expression context using two weakest-precondition predicates,
$\WPpexpr$ (Fig.~\ref{fig:wp-pexpr}) for place expressions, and
$\WPvexpr$ (Fig.~\ref{fig:wp-vexpr}) for value expressions.

The predicate $\WPpexpr~e~\xi~\CFpost{\ell}{ Q }$ is the weakest precondition of an expression $e$ in function frame $\xi$,
for a \emph{control-flow sensitive} postcondition $\CFpost{\loc}{Q}$ that depends on location $\ell : \Loc$, the result of evaluating $e$.
Similarly, $\WPvexpr~e~\xi~\CFpost{v}{Q}$ is the weakest precondition of an expression $e$ in value expression context
for a postcondition that depends on value $v : \Val$, the result of evaluating $e$.
Note that the postcondition is determined by the \emph{ambient} expression, which also determines whether
the operand expression of $\WPx{}$ predicate appears in place or value context.

As a simple example, the weakest precondition of a boolean literal
in value context, $\WPvexpr~\Rtrue~\xi~\CFpost{v}{Q}$, reduces via
\WPVexprLiteralBool{} directly to $Q~(\valBool{true})$: for any postcondition $Q$,
$\Rtrue$ satisfies it as long as $Q$ holds for the value representing boolean true.
Integer literals work similarly, via \WPVexprLiteralInt{}.

A path referring to a local variable, e.g.\ $\Rvar{foo}$, is a place expression;
obtaining $\WPpexpr~\Rvar{foo}~\xi$ amounts to asserting the variable is present
in the current frame $\xi$ and returning its location
(cf.\ \WPPexprVar). A local-variable path can also occur in value context.

\paragraph{Move and copy semantics}
\label{sec:movecopy}
When a place expression occurs in place context, or a value expression in value context,
the weakest precondition is accumulated directly.
When a place expression occurs in value context---e.g.\ in a pattern binding---it denotes
the value held at that location. If the value's type implements the \lstmath{Copy} trait, it is copied;
otherwise it may be \emph{moved} out, de-initializing the location.
We represent moving out of a location with the $\WPmove$ predicate:

\begin{definition}[Move semantics, $\Rhrefdoc{expressions.html\#r-expr.move}{\WPmove} : \Loc \to \WPx{\Val}$]
  \figuresize
  $$
    \WPmove ~ \loc ~ Q \triangleq \exists v . \loc \pointsto v \ast (
    \loc \pointstoUninit \toRTy{v}) \mwand Q~v
  $$
\end{definition}
\noindent
When moving out of a location $\loc$, the postcondition $Q$ is enriched with a separation logic obligation
of ownership of a value $v$ represented at the location, $\exists v. \loc \pointsto v$.
The original
postcondition is provided an assumption that location $l$ has been uninitialised. Note that the
encoding logic is affine and thus, we can read the statement such that we are required to ``exchange'' the
$\loc \pointsto v$ fact for $\loc \pointstoUninit \toRTy{v}$, i.e. move the value $v$ out of location $\loc$.

To continue with the example of the local variable $\Rvar{foo}$, variables
that are not currently borrowed can be moved out\footnote{cf. \cite[expr.move.movable-place]{rustref}}.
When $\Rvar{foo}$ occurs in value context, $\WPvexpr~\Rvar{foo}~\xi~Q$ can be constructed
using $\WPVexprVar$ by sequentially first constructing the weakest precondition of the path expression
in place context and then moving the variable's value out of the place. The definition of $\WPmove$ ensures that
the variable is not currently borrowed using the ownership obligation.

A value is copied instead of moving when its type implements the \lstmath{Copy} trait.
At the level of semantics, we treat
copy as an optimization of a move: moving a value when copy was correct uninitializes the location, resulting in an unsatisfiable goal if the value is accessed again.
This allows generic functions to be verified once under move semantics, with proofs reused for Copy-type monomorphizations. 
\looseness=-1 We refer the reader to the Rocq formalization for details.

\paragraph{Temporaries}
When a value expression is evaluated in place expression context,
a temporary unnamed memory location (\emph{a temporary}) is created.
It is initialized to the value, to which the expression then evaluates,
and dropped at the end of its drop scope, usually the end of the enclosing statement.
We define the appropriate predicate:

\begin{definition}[Semantics of temporaries, $\Rhrefdoc{expressions.html\#r-expr.temporary}{\WPtmp} : \Val \to \WPx{\Loc}$]
  \figuresize
  \[
  \WPtmp~v~Q \triangleq
  \forall \loc . \loc \pointsto v \mwand Q~(\loc : ds)~\loc
  \]
\end{definition}

\noindent
\noindent The $\WPtmp$-transformed postcondition in goal position means a universally qualified location $\loc$ can be introduced into context
as a fresh location, as well as the fact on the left side of the separation implication that the location now contains a representation of value $v$.

Building on our running example, consider the weakest precondition of borrowing a literal,
$\WPvexpr~\lstmath{&true}~\xi~\CFpost{v}{Q}$.
Using the inference rule \WPVexprBorrowValue{}, the weakest precondition is reduced to
$(v \leftarrow \WPvexpr~\lstmath{true}~\xi ;~ \WPtmp~v)~\CFpost{v}{Q}$ and,
after inlining the definition of $\WPtmp$, into
$\CFpost{\_}{\forall \loc, \loc \pointsto \valBool{true} \mwand Q~\valBool{true}}$,
that is, the postcondition holds for any fresh location $\loc$
that stores the boolean value true.

The fresh location $\loc$ introduced above is itself a temporary, and is dropped once its
enclosing scope ends; the full drop scope mechanism---including statement temporaries,
block-scoped locals, and arm bindings---is described in Appendix~\ref{apx:dropscopes}.

\paragraph{Arithmetic and logical operators}
Arithmetic and logical operators allow for largely uniform treatment.
In particular, unary negation (\WPVexprUnOpNeg{}) is obtained by sequencing the weakest precondition of an operand with the following predicate:

\begin{definition}[Semantics of unary negation,
    $\Rhrefdoc{operator-expr.html\#r-expr.negate.results}{\WPuneg} : \mathit{UOp} \to \Val \to \WPx{\Val}$]
  \figuresize
  \begin{align*}
\WPuneg~\lstmath{!}~(\valBool{b})~Q & \triangleq Q (\valBool{(\neg b)}) \\
      \WPuneg~\lstmath{-}~(\valInt{s}{w}{z})~Q & \triangleq \noover s~w~(\text{-}z) \ast Q~(\valInt{s}{w}{(\text{-} z)})
  \end{align*}
\end{definition}

\noindent
The predicate $\WPuneg$ asserts the result of negation is \emph{representable} in the operand's type without overflow; this trivially holds for booleans and signed integers, and is unsatisfiable for unsigned ones.
Binary operators \WPVexprBinOp{} work similarly: the weakest preconditions of both operands are obtained, then an assertion that the result is representable; we omit the details of $\WPbinop$ and refer to the formal development.

\paragraph{Structured values}

\begin{figure}\RhrefToDocstrue
  \figuresize
  \begin{flushleft}
    \boxed{$$\WPpexpr : \Expr \to \Xi \to  \WPx{\Loc)}$$}
  \end{flushleft}

  \newbox\bTuple
  \setbox\bTuple\hbox{\Rtuple{es}}
  \newbox\bTupleField
  \setbox\bTupleField\hbox{e\lstmath{.}n}
  \newbox\bAdt
  \setbox\bAdt\hbox{$(\Radt{d}~fs)$}
  \newbox\bAdtField
  \setbox\bAdtField\hbox{e\lstmath{.}d}
  \newbox\bBlock
  \setbox\bBlock\hbox{$\Rblocke{ss}{e^?}$}
  \newbox\bIf
  \setbox\bIf\hbox{$\Rif~e_c~e_t~(\Relse~e_f)^?$}
  \newbox\bMatch
  \setbox\bMatch\hbox{$\Rmatch{e_\text{scrt}}{pes}$}
  \newbox\bLoop
  \setbox\bLoop\hbox{$\Rloop{\Rblocke{ss}{e}}$}
  \newbox\bBreak
  \setbox\bBreak\hbox{$\lstmath{break}$}
  \newbox\bContinue
  \setbox\bContinue\hbox{$\lstmath{continue}$}
  \newbox\bCall
  \setbox\bCall\hbox{$\Rfncall{fn}{es_\text{args}}$}
  \newbox\bReturn
  \setbox\bReturn\hbox{$\Rreturn{e}$}

  \begin{mathpar}
    \mprset{flushleft}
    \inferrule[\WPVexprTuple]
              {vs \leftarrow \WPvexprs~es~\xi ;
              \\\\
              \text{let}~ fs = [ (i, v) \mid i \mapsto v \in vs ] ~\text{in}
              \\\\
              \quad\retCF \valTuple{fs}
              }
{\WPvexpr~\usebox\bTuple{}~\xi}

    \inferrule[\WPVexprAdt]
              { vs \leftarrow \WPvexprs~fs.2~\xi ;
                \\\\
               \text{let}~ fs = [ (d, v) \mid d \mapsto v \in \zip fs.1~vs  ] ~\text{in}
               \\\\
              \text{let}~ l = \mathtt{layout}~d ~\text{in}
              \\\\
              \quad\retCF \valAdt{l}{fs}
              }
{\WPvexpr~\usebox\bAdt{}~\xi}

    \inferrule[\WPPexprTupleField]
              { \loc \leftarrow \WPpexpr~e~\xi~;
                \\\\
                v \leftarrow \WPvalOver~\loc~;
                \\\\
                \text{if}~\lookupTuple{v}{n}~\text{then}
                \\\\
                \quad \retCF \loc.n
                \\\\
                \text{else}~\bot
              }
{\WPpexpr~\usebox\bTupleField{}~\xi}

    \inferrule[\WPPexprAdtField]
              { \loc \leftarrow \WPpexpr~e~\xi~;
                \\\\
                v \leftarrow \WPvalOver~\loc~;
                \\\\
                \text{if}~\lookupAdt{v}{n}~\text{then}
                \\\\
                \quad \retCF \loc.d
                \\\\
                \text{else}~\bot
              }
{\WPpexpr~\usebox\bAdtField{}~\xi}

  \inferrule[\WPVexprIf]
            {v \leftarrow \WPvexpr~e_c~\xi~;
              \\\\
              \text{if}~v~\text{is}~true~\text{then}~
\WPvexpr~e_t~\xi
              \\\\
              \text{else if}~e_f~\text{then}~
\WPvexpr~e_f~\xi
              \\\\
              \text{else}~\retCF~\valUnit
            }
            {\WPvexpr~\usebox\bIf{}~\xi}

   \inferrule[\WPVexprBlock]
             {\WPblock~ss~e~\xi}
             {\WPvexpr~\usebox\bBlock{}~\xi}

   \inferrule[\WPVexprMatchV]
             {v \leftarrow \WPvexpr~e_\text{scrt}~\xi ~;
               \\\\
               \WParms~e_\text{scrt}~pes~\xi~;
               \\\\
               \guard \MatchBreak
               \and \quad e_\text{scrt}~\text{is vexpr}
             }
             {\WPvexpr~\usebox\bMatch{}~\xi}

   \inferrule[\WPVexprMatchP]
             {\loc \leftarrow \WPpexpr~e_\text{scrt}~\xi ~;
               \\\\
               v \leftarrow \WPvalOver~\loc
               \\\\
               \WParms~e_\text{scrt}~pes~\xi~;
               \\\\
               \guard \MatchBreak
               \and \quad e_\text{scrt}~\text{is pexpr}
             }
             {\WPvexpr~\usebox\bMatch{}~\xi}

   \inferrule[\WPVexprLoop]
             {\WPblock~ss~e~\xi ~;
               \\\\
               \guard~\Continue
               \\\\
               \WPvexpr~\usebox\bLoop{}~\xi
               \\\\
               \guard~\Break
             }
             {\WPvexpr~\usebox\bLoop{}~\xi}

   \inferrule[\WPVexprBreak]
             {\throw~\Break
             }
             {\WPvexpr~\usebox\bBreak{}~\xi}

   \inferrule[\WPVexprContinue]
             {\throw~\Continue
             }
             {\WPvexpr~\usebox\bContinue{}~\xi}

   \inferrule[\WPVexprCallGlobname]
             {vs \leftarrow \WPvexprs~es_\text{args}~\xi ~;
               \\\\
               \WPfn~fn~crate~vs
             }
             {\WPvexpr~\usebox\bCall{}~\xi}

   \inferrule[\WPVexprReturn]
             {v\gets \WPvexpr~e~\xi ~;
               \\\\
              \throw\;(\Return~v)
             }
             {\WPvexpr~\usebox\bReturn{}~\xi}

  \end{mathpar}
  \ff{move \textsc{WP-*-ADT*} to appendix if space needed}
\caption{Weakest precondition predicates for structured values}
  \label{fig:wp-vexpr2}
\end{figure}\RhrefToDocsfalse

Structured values in Rust are heterogeneous tuples, homogeneous, fixed-size arrays, and user-defined ADTs, structs and enums.
The semantics of creating a value of tuple type is captured by \WPVexprTuple{}.
First, the weakest precondition of expressions in fields positions are collected via $\WPvexprs$ and then a finite map from field index to field value is constructed using enumerated list comprehension.
The semantics of $\WPvexprs$ corresponds to sequentially collecting the weakest preconditions of individual expressions over the list of expression values:

\begin{definition}[Semantics of value expression sequences, $\WPvexprs$]
\figuresize
\begin{align*}
  \WPvexprs~\nil~\xi \triangleq~ & \retCF \nil \\
  \WPvexprs~(e \cons es)~\xi \triangleq~ & v \leftarrow \WPvexpr~e~\xi~; \\
  & vs \leftarrow \WPvexprs~es~\xi~; \\
  & \retCF (v \cons vs)
\end{align*}
\end{definition}
\noindent
Consider, e.g., $\WPvexpr \lstmath{(True, 42u8)} \{v . Q~v \}$. Using the rule
\WPVexprTuple{} and the rules for literals, we obtain
$Q~(\valTuple{(0, \valBool true), (1, \valInt{u}{8}{42})})$.
Indexing into tuples is captured by \WPPexprTupleField{}.
Indexing expressions are place expressions, and $\WPpexpr$ of the operand expression is obtained, and then the precondition is enriched
with an assertion that the location contains a value:

\begin{definition}[Value-at-location, $\WPvalOver : \Loc \to \WPx{\Val}$]
\figuresize
\begin{align*}
  \WPvalOver~\loc~Q \triangleq
  \exists v . \loc \pointsto v \ast
                           (\loc \pointsto v \ast \wand Q v)
\end{align*}
\end{definition}
\noindent
Overall, if the value map is defined at the index ($\lookupTuple{v}{n}$),
the precondition is bound to the location of the tuple $\loc$ offset by the index,
and it is unsatisfiable otherwise.
ADTs are treated in a corresponding way, but are modeled by finite maps from field names to values, and the type identifier is used to decide layout of the value. We refer to the Rocq development for details.

\paragraph{Block expressions}
Block expressions
serve as declaration scopes and as control-flow expressions.
As a control flow expression, a block comprises a sequence of statements and an optional final expression, evaluated sequentially.
A block is a value expression evaluating to the value of the final expression, or to unit if there is none.
Its semantics is captured by the following predicate:

\begin{definition}[Semantics of blocks, $\Rhrefdoc{expressions/block-expr.html\#r-expr.block.evaluation}{\WPblock} : \text{list}~\Stmt \to \Expr \to \Xi \to \WPx{\Val}$]
  \label{def:wp-block}
\figuresize
\begin{align*}
  \WPblock ~(s \cons ss)~e^?~\xi \triangleq~&
  \xi \leftarrow \WPstmt~s~\xi ~;
\WPblock~ss~e^?~\xi \\
  \WPblock~\nil~e~\xi \triangleq~& \WPvexpr~e~\xi \\
  \WPblock ~\nil~\emptyset~\xi \triangleq~& \retCF~\valUnit
\end{align*}
\end{definition}
\ff{line compressed}

\noindent
For a block with a leading statement, the weakest precondition of the first statement is accumulated, then those of the remaining statements; a final expression, if present, contributes its own, with the postcondition applied to its value (or unit otherwise).
The rule $\WPVexprBlock{}$ in $\WPvexpr$ then reduces a block's weakest precondition to $\WPblock$; e.g.\
$\WPvexpr~\Rblocke{\lstmath{()}}{42}~\xi~Q$ becomes
$(\WPstmt~\lstmath{()}~\xi~; \WPvexpr~\lstmath{42}~\xi)~Q$.

\paragraph{If expressions}

The if expression is the simplest non-trivial case of control flow: a boolean condition operand, a block, and optionally an else block.
The condition is evaluated; if $\texttt{true}$, the block is evaluated, otherwise the else block if present.
It is a value expression yielding the value of the evaluated block, or $\lstmath{()}$ if none was.
The semantics is captured by \WPVexprIf{}. In this simple form, applying the inference rule duplicates the postcondition;
we address this in Section~\ref{sec:proglog} via join-point reasoning.

\paragraph{Match expressions}

A match expression sequentially compares the value of a \emph{scrutinee} to the
patterns of its \emph{arms}, evaluating the expression of the first arm whose pattern
matches. It is the first construct in this paper that combines pattern-match failure
($\PatNoMatch$) with control-flow-transferring constructs ($\MatchBreak$) in the control flow monad.
The weakest precondition of an arm is:

\begin{definition}[Semantics of match arms,
    $\Rhrefdoc{expressions/match-expr.html\#r-expr.match.scrutinee-value}{\WParms}
    : \Val \to \text{list}~\Pat \to \Xi$ and $\WParm$]
\figuresize
\begin{align*}
  \WParm~v_\text{scrt}~(\Rarm{pat}{e})~\xi \triangleq~&
  \WPpat~pat~e_\text{scrt}~\xi ~;
\WPvexpr~e~\xi ~;~
  \MatchBreak
  \\
\\
  \WParms~v_\text{scrt}~(\Rarms{a}{as})~\xi \triangleq~&
  \WParm~a~\xi ~;
\guard \PatNoMatch ~;
\WParms~v_\text{scrt}~as~\xi \\
  \WParms~v_\text{scrt}~\nil~\xi \triangleq~& \bot \\
\end{align*}
\end{definition}
The weakest preconditions of arms are collected in order.
If an arms control flow results in $\PatNoMatch$, this state is turned
back to sequential by the $\guard$ and the accumulation continues with
the remaining arms. The weakest precondition of an empty set of arms is
unsatisfiable since patterns in a match expression are exhaustive.
The precondition of an arm, $\WParm$, is that of the arm's pattern, sequentially
followed by that of the arm's expression and then the $\MatchBreak$ operation.
Recall that $\WPpat$ (Fig.~\ref{fig:wp-pat}) yields $\PatNoMatch$ on failed
matches, skipping the expression's precondition, and is guarded by the ambient $\WParms$;
similarly, $\WPvexpr$ may yield other control flow state, propagated outside $\WParms$.
\looseness=-1 A match expression semantics slightly differs depending whether
the scrutinee is a value or a place expression and is captured by
\WPVexprMatchV{} and \WPVexprMatchP{} respectively.

\paragraph{Loop, break, and continue expressions}
A loop expression repeats evaluation of its body.
A \Rbreak{} in the body immediately terminates the loop; a \Rcontinue{} terminates the current iteration.
We model loops in \WPVexprLoop{} by 1-unfolding of the loop body.
The unfolded body's precondition is guarded by $\Continue$, which
sequences the next iteration's precondition when evaluation was terminated by $\Rcontinue$.
The whole unfolding is guarded by $\Rbreak$, which sequences the precondition
of the program following the loop when evaluation terminated by $\Rbreak$. The semantics currently supports neither labels on break/continue (Def.~\ref{def:CF}) nor break values, though both should be straightforward to add; treating loops as 1-unfoldings requires user-supplied invariants.

\subsection{Function Items, Function Specifications, and Return Expression}
\label{sec:sem-fn}

Functions represent the key evaluation abstraction mechanism in Rust.
A crate consists of items, among them functions with given parameters.
Functions are called by call expressions that bind parameters to concrete arguments,
and a value is passed to a caller via a return expression.
Functions are moreover the primary unit of code at which we state specifications.
The semantics of functions is given as follows:
\begin{definition}[Semantics of functions,
    $\Rhrefdoc{items/functions.html\#r-items.fn.intro}{\WPfn}
    : \mathcal{D} \to \text{Crate} \to \text{list}~\Val$]
 \label{def:wp-fn}
\figuresize
\begin{align*}
\WPfn~fn~crate~vs_\text{args}~\triangleq~&
  \text{if}~crate[fn]~\text{is}~(pats, e_\text{fn})~\text{then}
  \\ &
\quad \xi \leftarrow \WPpats~pats~vs_\text{args}~\nil~;
  \\ &
  \quad \WPvexpr~(\Rreturn{e_\text{fn}})~\xi~;
  \\ &
  \quad \guard \Return
  \\&
  \text{else}~\bot
\end{align*}

\end{definition}
\noindent
At the level of AST, syntax is fully qualified. The fully-qualified name of a function is used to look up parameters and body of the function in its defining crate.
Function parameters are patterns, and the precondition of matching the patterns against the real arguments is accumulated via $\WPpats~pats~vs_\text{args}~\nil$.
The resulting initial frame $\xi$ is used to obtain the precondition of the body of the function, $\WPvexpr~(\Rreturn{e_\text{fn}})~\xi$.
Note that the body obtained from the crate is administratively
wrapped in a return expression\footnote{cf. \cite[items.fn.body.intro]{rustref}}.
Accumulation of the weakest precondition of the function thus always results in $\Return{}$ and the guard always succeeds.
The semantics of return (\WPVexprReturn{}) accumulates the
precondition of the operand and binds its value to the
$\Return$ operation. This value is then caught by the guard in $\WPfn$ and
is the function's return value.

Function call semantics (\WPVexprCallGlobname{}) amounts to accumulating
the precondition of evaluating function arguments, in order,
and then the precondition of the function itself, $\WPfn$,
bound to values of arguments.
The semantics of inherent and trait implementations works similarly,
at the level of the AST the headers and call sites of these and of ordinary
functions are uniform and, the only difference is the increased amount of book-keeping. We refer to the Rocq development for details.

The above reasoning principles allow
\emph{inlining} the proof of a callee into a proof of its caller.
To modularize reasoning about caller and callee,
and to reason about function items in a crate,
we need a notion of function \emph{specification}
and means of connecting a specification to semantics of functions.

\begin{definition}[Specifications of functions, $\Spec$ and $\FnItemOk$]
  \label{def:fn-item-ok}
  \figuresize
  \begin{align*}
    \Spec \triangleq \forall (args : \text{list}~\Val) \to
       (\forall (r : \Val) \to \iProp) \to \iProp
  \end{align*}
  \begin{align*}
    \FnItemOk~fn~crate~spec \triangleq
    \square (\forall Q~vs_\text{args} . spec~vs_\text{args}~Q \mwand
    \WPfn~fn~vs_\text{args}~Q)
  \end{align*}
\end{definition}
\noindent
Function specifications are proposition formers in $\iProp$ taking
the list of argument values, and a postcondition in $\iProp$ over
the return value of the function.
The persistence modality ($\square$) marks a resource as freely duplicable,
so that a proven specification can be reused at every call site.
Proving a function $fn$ in a crate correct with respect to a
specification--expressed as  $\FnItemOk$--means
to show, under $\square$, that for any postcondition $Q$ and
argument values, the specification entails the weakest precondition
of the function over $Q$.

Modular reasoning about a caller and callees is achieved by proving that the correctness of a caller ($\FnItemOk$) is established from correctness assumptions about callees.
 \section{Proving Soundness Modularly with ITrees}\label{sec:itreemodel}

The classical approach to defining an operational semantics of a language
is \emph{monolithic}: execution of each and every language construct
is described in terms of changes to whole-program state.
Interaction trees -- \itrees -- provide a framework for defining modular, compositional semantics
of effectful programs in Rocq.
Recent work by~\citet{plalacarte} delivers a framework for
deriving a weakest precondition
calculus from an \itree semantics. 
In this section, we illustrate
how we instantiate parameters $\WPvexpr$ and $\WPstmt$
(Def.~\ref{def:corepreds}) and how we
prove the soundness of the axioms
from Fig.~\ref{fig:wp-vexpr} using this approach.
Due to space limitations, we refer the reader to~\citet{itrees} and
\citet{plalacarte} as well as to Appendix~\ref{apx:itree} for more detail on
the underlying framework.

We start from a subset of interest, here,
the pure binary arithmetic operations.
The evaluation of an arithmetic expression $e$ either results in an integer
or in undefined behavior due to an arithmetic overflow error. The semantics of $e$ is
therefore given by its denotation $\itreeof{e} : \itreeER{\ubE}{\mathcal{V}}$,
an \itree with return values in $\mathcal{V}$ and events in $\ubE$,
modeling undefined behavior.
Following~\citet{plalacarte}, we derive a weakest precondition
calculus for $e$ from its \itree denotation.
Intuitively, the weakest precondition for $e$ to satisfy a predicate
over values $\Phi$ is that the execution of $e$ results in a value
satisfying $\Phi$. If, instead, the execution of $e$ leads to an
integer overflow, $\Phi$ can't be satisfied, so that the weakest
pre-condition for triggering a \ubE\; is set to $\bot.$
More concretely, based on the \itrees semantics of
binary operations we prove that the result of adding two operands
is equal to the mathematical sum of the operand values as long as
there is no arithmetic overflow:
\newbox\bAdd
\setbox\bAdd\hbox{$e_l~\lstmath{+}~e_r$}
\begin{mathpar}
  \figuresize
  \inferrule[Binop-add]
            {\WPvexpr~e_l~\xi ~\{ r . r = v_l \}\ast
\WPvexpr~e_r~\xi ~ \{ r. r = v_r \}\ast
\noover~(v_l + v_r)
            }
            {\WPvexpr~\usebox\bAdd{}~\xi~\{ v . v = v_l + v_r \}}
\end{mathpar}
We observe that \textsc{Binop-add} is an instance of the axiom
\WPVexprBinOp{} from~\cref{fig:wp-vexpr}. The
proof of \textsc{Binop-add} readily adapts to \WPVexprBinOp{}, showing it
holds relative to the \itrees semantics for binary operations.

Having established the approach on a pure fragment,
we now extend the \itree semantics to a larger class of expressions
that includes heap effects, e.g. \lstmath{let mut x = v}.
The semantics of an assignment is an
\itreeER{(\texttt{heapE}\;+'\;\ubE)}{\mathcal{V}}, modeling a
computation that can either result in a value, trigger an overflow
\ubE, or have an effect on the heap, described by the event type
\texttt{heapE}. Intuitively, a \texttt{heapE} corresponds to \emph{get} or \emph{set}
the value at a certain location $l$.  For the triggering of
\texttt{heapE} to satisfy a certain predicate $\Phi,$ it is necessary
to prove ownership of the location $l$ and that the current or the
updated value satisfy $\Phi$.
As a consequence, the weakest pre-condition for \lstmath{let mut x =
  v} to satisfy $\Phi$ requires $x$ to point to a location $l$ that,
if updated with the value $v$, guarantees $\Phi$:
\newbox\bLet
\setbox\bLet\hbox{\lstmath{let mut x = v}}
\begin{mathpar}
  \inferrule[\textsc{LetV-lemma}]
      { (x \mapsto l)\;\ast\; ((l \mapsto v) \wand\; \Phi ) }
      { \WPstmt\;(\usebox\bLet)~\xi~\{ \Phi \} }
\end{mathpar}

\noindent
The modularity of this approach now becomes crucial. The \itrees
library allows to naturally lift an $\itreeER{\ubE}{\mathcal{V}}$ into
an $\itreeER{(\texttt{heapE}\;+'\;\ubE)}{\mathcal{V}}$ and, at the
cost of proving an adequacy result~(cf. \cite{plalacarte}), to soundly use
the weakest precondition rules proven for binary operation when
reasoning about programs involving assignments. We can thus prove:
\newbox\bLetB
\setbox\bLetB\hbox{$\lstmath{let mut x =}~ (v_l + v_r)$}
\begin{mathpar}
  \inferrule[Let-addV-lemma]
      { (x \mapsto l)\;\ast\;
               \noover~(v_l + v_r)\;\ast\;
               (v = v_l + v_r)\; \ast\;
               ((l \mapsto v) \wand\; \Phi) }
             { \WPstmt\;(\usebox\bLetB{})~\xi~\{ \Phi \} }

\end{mathpar}

\noindent
We observe that both \textsc{LetV-lemma} and \textsc{Let-addV-lemma}
are instances of \WPStmtExpr\; from~\cref{fig:wp-stmt}, which can be
proven by modularly extending the \itrees semantics to all $s \in
\Stmt$ following the approach illustrated above.

To conclude, we establish the soundness of the axioms from Sec.~\ref{sec:sem}
incrementally, by modularly extending the \itrees semantics to further terms
in $\Expr$ and $\Stmt$. Where necessary, we introduce new event types,
e.g. \texttt{heapE} above, and show the reasoning
principles for the events introduced are \emph{adequate} with respect
to the reasoning principles established for the other event types.
Completed proofs are final: \itrees's compositionality guarantees they
remain valid as coverage grows, so lifting, e.g., an $\itreeER{\ubE}{\cdot}$
denotation into $\itreeER{(\texttt{heapE}\;+'\;\ubE)}{\cdot}$ costs one
adequacy lemma for the new event type, never a re-proof of the existing
fragment. We have completed this argument for the pure fragment and
heap-manipulating statements illustrated above; per-axiom status for the
remaining constructs is itemized in Appendix~\ref{apx:soundness}.
However, full soundness establishes only internal consistency--
the axioms are satisfied by an operational model--
and does not guarantee that the axioms capture the intended semantics
of Rust as a programmer would expect;
test-suite validation we discuss in Sec.~\ref{sec:sem} is therefore a
necessary complement.

 \section{Program Logic}\label{sec:proglog}

The axiomatic semantics in Sec.~\ref{sec:sem} provides a basis for reasoning about program behavior.
However, it operates at a low level of abstraction, making direct proofs detailed and unwieldy.
Moreover, the direct encoding of Rust semantics leads to large proof goals in Rocq, which
slow down both interactive proof development and final proof checking (\emph{Qed-time}) in Rocq.
To address these issues, we develop an additional reasoning layer in the form of a program logic.
As motivated in Sec.~\ref{sec:motivation}, proofs need to be structured in a way that the axiomatic semantics does not provide:
(1) Controlling WP accumulation through sealing, which improves readability, controls the evaluation order to limit proof term size and improves performance.(2) Reordering WP accumulation, which enables reasoning in an idiomatic manner, closely aligned with the structure of the Rust source code.
(3) Join-point reasoning supporting modular reasoning about shared proof contexts.

\paragraph{Directing evaluation}\label{subsec:direval}

\begin{figure}
  \figuresize

  \newbox\bBlockCons
  \setbox\bBlockCons\hbox{\Rblocke{s~\lstmath{;}~ss}{$e^?$}}
  \newbox\bBlockConsA
  \setbox\bBlockConsA\hbox{\Rblocke{ss}{$e^?$}}
  \newbox\bIfElse
  \setbox\bIfElse\hbox{$\Rif~e~e_T~\Relse~e_F$}

  \begin{mathpar}
    \mprset{flushleft}
    \inferrule[\WPBlockCons]
              {\xi' \leftarrow \WPstmt~s~\xi~;
                \\\\
                \WPblockSealed~\usebox\bBlockConsA{}~\xi' }
{\WPstmt~\usebox\bBlockCons{}~\xi}

              \inferrule[\WPIfJoin]
                        { ~ \\
                          \WPvexpr~e_T~\xi~(P~true) \mwand R
                          \\
                          \\\\
                          \WPvexpr~e ~\xi~(\lambda v. P v \ast F)
                          \\
                          \WPvexpr~e_F~\xi~(P~false) \mwand R
                          \\
                          R \ast F \mwand Q
                        }
                        {\WPvexpr~(\usebox\bIfElse{})~\xi~Q}

  \end{mathpar}

  \caption{Program logic, derived lemmata (full set, including
    \WPBlockNilExpr, \WPBlockNilNoExpr, and \WPPatConst, in
    Appendix~\ref{apx:proglog})}
  \label{fig:proglog}
\end{figure}

As discussed in Section~\ref{sec:sem}, $\WPblock$--the weakest precondition of a block expression--
is defined recursively as a fixpoint over the block’s sequence of statements:
the $\mathrm{wp}_\text{block}$ definition
for a block expression containing many statements
unfolds into a long, nested chain of sequentialized statements.
To mitigate this issue, we employ a standard technique in Rocq: \emph{sealing} definitions,
that is, making them opaque and exposing only specific lemmata for controlled unfolding.
In the case of \WPblockSealed, we provide three such lemmata (Fig.~\ref{fig:proglog}):
\WPBlockCons{} for blocks with a leading statement,
and \WPBlockNilExpr{} and \WPBlockNilNoExpr{} for empty blocks, depending on whether a trailing expression is present.
The first rule enables evaluating the leading statement first, keeping the
remaining block opaque until needed.\footnote{Appendix~\ref{apx:proglog} gives the full walkthrough.}
Note that these lemmata, and all others in the program logic, are proven from
the axioms of the semantics and are thus sound relative to it.

\paragraph{Reordering evaluation}\label{subsec:reval}

WP accumulation in the semantics proceeds by structural traversal of the AST,
which does not always align with the order in which a programmer would
naturally reason about evaluation---for instance, matching against a literal
pattern accumulates the WP of the scrutinee before that of the pattern's
constant, though the constant's evaluation does not depend on the scrutinee.
We introduce a sealed definition $\delayedCpat$ that reorders this: it
accumulates the WP of the pattern's constant expression first, then that of
the scrutinee, checking the match once both are available. This reordering is
sound because the two evaluations are independent---there are no side
effects or dependencies that would constrain their order.\footnote{Appendix~\ref{apx:proglog}
gives $\delayedCpat$'s definition and the derivation in full.}

\paragraph{Join-point reasoning}\label{subsec:joinpoint}

A third situation in which the program logic substantially simplifies proofs arises with join-point reasoning.
Several language constructs evaluate a scrutinee and dispatch control flow based on the result; the simplest instance is the if expression.
When accumulating WPs, the value of the test expression may not be concrete,
requiring both branches to be analyzed independently, and the remainder of
the program after the conditional is often identical regardless of which
branch was taken, leading to duplicated proof and inflated proof terms.
To avoid this duplication, we employ join-point reasoning, captured by the
inference rule \WPIfJoin{} (Fig.~\ref{fig:proglog}): we identify a common
precondition $P$ sufficient for either branch to establish a shared
intermediate postcondition $R$, discharge the branches independently against
$R$, and prove the shared continuation once, composing $R$ with the frame
$F$--the spatial resources not consumed by the branch proofs--to establish
the overall postcondition $Q$.
This technique is not limited to if expressions: the same style of reasoning
applies to \lstmath{match} expressions, and reasoning about \lstmath{loop}
expressions also requires join-point reasoning, with the loop invariant
serving as the join-point pre- and post-condition connecting iterations.\footnote{Appendix~\ref{apx:proglog} walks through the general case.}

\paragraph{Defunctionalized continuation stacks}\label{subsec:defuncont}

Sealing~$\WPblockSealed$ controls proof goal size for straight-line code, but a deeper
source of blowup remains: the $\bindCF$ combinator accumulates the sequential
postcondition~$Q_\text{seq}$ as a nested closure at every step
($\bindCF~f~w~Q_\text{seq}~Q_\text{rt} = w~(\lambda a.~f~a~Q_\text{seq}~Q_\text{rt})~Q_\text{rt}$),
producing deeply nested closure trees in proof goals for long blocks.
Guard combinators (Sec.~\ref{sec:sem}) similarly rebuild both continuations at each
scope boundary.
These closure trees cause slow \textit{Qed}-times and make proof states
difficult to read.

We address this by \emph{defunctionalizing} the two continuations into sealed
first-order stacks:

\begin{definition}[Continuation stacks]
\figuresize
\begin{align*}
  \scont~A &::= \SKnil~(Q_\text{seq} : A \to \iProp) \mid \SKbind~(f : A \to \WPx{B})~(\mathit{rest} : \scont~B) \\
  \hcont &::= \HKnil~(Q_\text{rt}) \mid \mathit{HK}_X~f~(\mathit{S} : \scont)~\mathit{rest}
\end{align*}
\end{definition}

\noindent
$\scont~A$ is a type-aligned list of pending Kleisli arrows---each $\SKbind$ frame
corresponds to one application of $\bindCF$.
$\hcont$ is a list of installed guard scopes; $\mathit{HK}_X$ entries are pushed by
$\mathit{guard}_{X\text{-CF}}$ combinators at loop, function, and match boundaries.
The key algebraic fact is that $\bindCF$ threads~$Q_\text{rt}$ unchanged---only the
guard combinators install a new control-flow handler---so the handler stack is
\emph{shared} across the entire $\bindCF$ chain: a proof goal contains one $\hcont$,
not one per bind step. Branching constructs (\lstmath{if}, \lstmath{match}) do not
break this sharing: as detailed below, join-point reasoning lets both branches
continue into the same $\scont$ tail, so the shared continuation is not duplicated
per branch either.

The two stacks are given sealed denotation functions $\seqK$ and $\rtK$:
\begin{align*}
  \seqK~Q_\text{rt}~(\SKnil~Q) &= Q \\
  \seqK~Q_\text{rt}~(\SKbind~f~S) &= \lambda a.~f~a~(\seqK~Q_\text{rt}~S)~Q_\text{rt} \\
  \rtK~(\HKnil~Q_\text{rt}) &= Q_\text{rt} \\
  \rtK~(\mathit{HK}_X~f~S~H) &= \mathit{guard}_X~f~(\seqK~(\rtK~H)~S)~(\rtK~H)
\end{align*}
Sealing both functions with \texttt{mlock} prevents \texttt{cbn}/\texttt{simpl}
from expanding the stacks back into closure trees.
The equational lemmas $\seqK\text{-unfold}$ and $\rtK\text{-unfold}$
unfold \emph{exactly one frame} per rewrite, driving the proof forward without
exposing the full continuation.
Each $\mathit{guard}_{X\text{-CF}}$ scope installation pushes one $\mathit{HK}_X$
handler and one $\SKbind$ frame; each $\bindCF$ step pushes one $\SKbind$ frame.
Proof-mode tactics ($\mathit{tac\_bind\_norm}$, $\mathit{tac\_guard\_X}$, etc.)
are re-proved against $\seqK~(\rtK~H)~S$ / $\rtK~H$ and keep their existing names,
so existing proof scripts require no changes beyond the renamed continuation arguments.

Sequential join-points share~$\scont$ tails structurally (persistent linked lists),
and $\rtK~H$ is let-bound once per scope so it is not syntactically duplicated in the
proof goal.
The result is a proof state that carries a compact pair $(S, H)$ of first-order
data structures in place of a deeply nested closure tree, substantially reducing
\textit{Qed}-times and improving readability of intermediate goals.
 \section{Automation}
\label{sec:auto}

\ff{by symbolic stepping the Iris logic in Coq metalogic}
\ff{describe the stages (wp, CF, sep + arith)}
\ff{briefly describe the outlook, potential for more meta-programmed, modular approach with
custom automation theories (elpi)}
\ff{describe how things fit compared to =, e.g. Ike's thesis, caesium, etc.}

\ff{add that:
  This design facilitates direct correspondence between proofs in our
framework and the language’s intended behavior as perceived by system
programmers and compiler engineers.  \ff{todo: move the below para to
  logic layer/intro/somewhere else?}  Building on this interface, we
support \emph{symbolic stepping} proofs, where execution of Rust
programs is simulated within the logic through symbolic evaluation of
program states \cite{FarkaAKE25plos}.
}

The inference rules of our program logic are syntax-directed and, importantly, can be formulated
such that their conclusions are non-overlapping.
This design choice directly enables proof automation:
for any given sub-goal with a WP predicate in head position, the appropriate rule can be selected deterministically,
allowing WP accumulation to proceed without manual rule selection.
We develop an automation layer that drives proofs by repeatedly applying the program logic rules.
Our automation pipeline proceeds through the following stages:

\begin{enumerate}
  \item \emph{Premise introduction and normalisation}: preprocessing the goal, introducing required hypotheses, and normalising its structure to expose the next reducible program construct.
  \item \emph{Logical inference rule selection}: deterministically choosing and applying the program-logic rule corresponding to the syntactic form of the current expression or statement.
  \item \emph{Sequentialisation of control flow}: restructuring the goal to reflect the sequential evaluation order imposed by the specification monad.\item \emph{Solving separation-logic side obligations}: discharging the spatial side conditions generated by the logic,
    such as ownership splitting, joining, or framing.
\end{enumerate}

This pipeline repeats until no further progress can be made,
performing WP accumulation automatically for large fragments of the program---the resulting tactic we call \texttt{wp\_engage}.
We also provide specialized tactics, e.g.\
\texttt{wp\_fn} for function prelude book-keeping,
\texttt{wp\_if} for join-point reasoning over if expressions,
and \texttt{wp\_call} for function calls;
factoring these out of the main loop
significantly improves the readability and maintainability of proof scripts~\cite{FMbedrock}.

Program-logic lemmata are not applied directly to entailments of the underlying separation logic;
instead they are lifted into Iris Proof Mode (IPM)~\cite{KrebbersJ0TKTCD18} entailments
by generic transformers, dispatched on the term under the WP by a dedicated Rocq
tactic. Whenever automation stalls, the full expressiveness of IPM remains
available to continue the proof manually.
\looseness=-1 The proofmode tactics operate directly on the defunctionalized continuation stacks
(Sec.~\ref{subsec:defuncont}), so that e.g.\ a $\bindCF$ application pushes a new $\SKbind$ frame
via $\seqK\text{-unfold}$ instead of building a closure.\footnote{Appendix~\ref{apx:proglog}
details the tactic-level mechanics.}

A distinguishing feature of THIR-level proofs is that the Rocq proof goal displays
the THIR AST under the WP, which pretty-prints directly to surface Rust---e.g.\ a \sys{}
goal for \lstmath{x + 1} looks like
$\WPvexpr~\lstmath{x + 1}~\xi~Q$. This contrasts with MIR-level tools, whose goals
expose an explicit basic-block transition with temporaries, forcing users to
reconstruct the correspondence to source code. This proximity to source, a
direct consequence of working at THIR, is the primary ergonomic benefit over MIR-level
foundational tools.

\paragraph{Evaluation}

Our automation already yields a substantial reduction in proof effort.
As noted in Sec.~\ref{sec:sem}, the test suite serves as semantic validation;
here we use it to compare proof effort per test, contrasting proofs carried out
directly via the semantics against automated ones.
Table \ref{tab:eval} reports non-blank lines of code
(measured by \texttt{cloc}\footnote{\url{https://github.com/AlDanial/cloc}}) for each test collection.

\setlength{\tabcolsep}{1.2em}
\begin{table}[h]
\figuresize
\centering
\begin{tabular}{lrrrr}
\toprule
\textbf{test} & \textbf{src} & \textbf{spec} & \textbf{raw} & \textbf{auto} \\
\midrule
arith     & 90  & 174 & 624  & 238 \\
arrays    & 25  & 78  & 222  & 106 \\
assign    & 23  & 48  & 268  & 48  \\
cf        & 24  & 41  & 88   & 43  \\
ifs       & 27  & 36  & 112  & 140 \\
impls     & 55  & 161 & 305  & 63  \\
matching  & 143 & 200 & 1073 & 142 \\
tuple     & 46  & 52  & 453  & 191 \\
types     & 56  & 51  & 59   & 30  \\
unary     & 18  & 32  & 112  & 39  \\
\bottomrule
\end{tabular}
\caption{Non-blank lines of code (src) and proofs in Rust source, and Rocq specification (spec), semantics based proofs (raw), and automated proofs (auto).}
\label{tab:eval}
\end{table}

\noindent
Overall, the automated proofs are roughly two to four times shorter than their raw counterparts.
Specifications are consistently about twice the size of the code, and
automated proofs about two to four times the size of the underlying Rust programs.
The anomaly for if expressions stems from the overhead of specifying join-point pre- and postconditions in a test file focused on them.
 \section{Case Study}
\label{sec:casestudy}

As a case study, we verify a buddy allocator~\cite{buddyalloc}, a standard
memory-management data structure in operating systems, implemented without
regard to ease of verification (approximately 300 lines of Rust, using
idiomatic features such as extensive match expressions). The top-level
specification of \Rfn{alloc} is, abbreviating the representation predicate
$\mathit{buddyRep}$ relating the Rust value to an abstract state $(bs, hs)$
of block-control and free-list-head lists---and folding in the free-list
well-formedness invariant over that abstract state---and writing
$\mathit{layoutRep}~l~\mathit{size}~\mathit{align}$ for the predicate relating
the caller-supplied \lstmath{Layout} argument $l$ to the requested size and
alignment:
\[
  \begin{gathered}
    \mathit{self} \pointsto v \ast \mathit{buddyRep}~v~bs~hs \ast \mathit{layoutRep}~l~\mathit{size}~\mathit{align} \ast (\text{bounds side-conditions on } \mathit{size}, \mathit{align})\\
    \vdash\; \WPfn~\Rfn{alloc}~[\mathit{self}, l]~\{v' \mid \exists bs'\,hs'.\ \mathit{self} \pointsto v' \ast \mathit{buddyRep}~v'~bs'~hs'\}
  \end{gathered}
\]
Allocation and deallocation are specified to preserve both the
representation and the free-list well-formedness invariant it carries, with
the concrete internal state existentially quantified in the postcondition.
The specification does \emph{not} claim functional correctness of the
allocation strategy---e.g.\ best-fit selection or any fragmentation
bound---only that whatever block is returned is disjoint from, and does not
corrupt, the allocator's remaining state; this is the appropriate notion of
correctness for a Rust allocator, which cannot rely on allocations actually
happening\footnote{cf.
\url{https://doc.rust-lang.org/alloc/alloc/trait.GlobalAlloc.html\#safety}}.
Establishing it requires functional specifications for all functions the
allocator calls---mostly showing that the free-block lists remain well-formed
after each operation---and is relative to (unproven) specifications of the Rust
\texttt{core} functions it uses. The resulting proof script,
including specifications, amounts to approximately 2800 lines of Rocq:
around 250 lines of function specifications, around 800 lines of general
reasoning about Rust constructs (including around 100 lines for join-point
pre- and postconditions), around 1400 lines of domain-specific lemmas
re-establishing the free-block-list invariants, and the remainder generic
Rocq lemmata. Automation and the program logic were instrumental in
keeping this size down; in particular, join-point reasoning
(Sec.~\ref{subsec:joinpoint}) eliminated otherwise necessary proof
duplication, leaving the Rust-specific reasoning mostly automated and the
bulk of the effort in domain-specific invariant re-establishment.

 \section{Related Work}\label{sec:relatedW}

\paragraph{Rust Verification}
Table~\ref{tab:toolcompact} compares \sys{} to the Rust verification tools discussed below.
\setlength{\tabcolsep}{0.5em}
\begin{table}[h]
\figuresize
\centering
\begin{tabular}{@{}p{0.24\linewidth}p{0.34\linewidth}p{0.08\linewidth}p{0.21\linewidth}@{}}
\toprule
\textbf{Tool} & \textbf{Level} & \textbf{Found.} & \textbf{Automation} \\
\midrule
Miri~\cite{miri} & MIR, dynamic & --- & Concrete execution \\
Kani~\cite{kani,kani2026} & MIR $\to$ Goto-C & No & SAT/SMT \\
Verus~\cite{verus} & Rust (ghost code) $\to$ SMT-LIB & No & SMT \\
Prusti~\cite{prusti} & MIR $\to$ Viper & No & SMT \\
RustHorn(Belt)~\cite{rusthornbelt} & Rust $\to$ CHC & No & CHC \\
Creusot~\cite{creusot} & MIR $\to$ Why3 & No & SMT \\
Flux~\cite{flux} & MIR & No & Highly automatic \\
Refined\-Rust~\cite{refinedrust} & MIR & Yes & Lithium \\
Aeneas~\cite{aeneas} & Rust $\to$ functional & Semi & Manual/interactive \\
\textbf{\sys{}} & \textbf{THIR} & \textbf{Yes} & \textbf{Syntax-driven} \\
\bottomrule
\end{tabular}
\caption{Rust verification tools comparison (cf.\ Table~\ref{tab:toolfull}, Appendix~\ref{apx:toolcomparison}, for a full comparison).}
\label{tab:toolcompact}
\end{table}
 Miri~\cite{miri} is the most widely used lightweight tool for detecting Rust memory-safety issues,
executing programs against the Rust Abstract Machine to catch undefined behavior in concrete
runs (Table~\ref{tab:toolcompact}). Kani~\cite{kani,kani2026} similarly targets Rust's MIR, but via
bounded model checking against CBMC's SAT/SMT backends rather than concrete execution.

Verus~\cite{verus} supports linear ghost types for separation-logic-style reasoning, and
Prusti~\cite{prusti} builds on the Viper~\cite{viper} verification infrastructure. RustHorn~\cite{rusthorn}
inspired subsequent Horn-clause-based work including Creusot~\cite{creusot}, which supports more
expressive specifications than Prusti via Why3~\cite{FilliatreP13}. Flux~\cite{flux} takes a
different, highly automatic approach via liquid (refinement) types with inferred invariants,
requiring far less annotation than Prusti or Creusot.
While these tools achieve a high degree of automation,
they all rely on external solvers,
which limits the expressiveness of specifications, precludes foundational end-to-end guarantees, and
introduces solvers into the TCB.
The tools most closely related to \sys{} are RefinedRust~\cite{refinedrust} and Aeneas~\cite{aeneas}.
RefinedRust, like \sys{}, is foundational, built on Iris, and supports modular reasoning,
providing a semantic type model for Rust in Rocq. Two differences are worth highlighting.
First, RefinedRust's MIR-level proof goals expose the control-flow graph, whereas \sys{}'s
THIR-level goals pretty-print directly to source Rust (Sec.~\ref{sec:auto} works through a
concrete example). Second, the theoretical encodings differ: \sys{} is based on Hoare-type
theory with a weakest-precondition specification monad (Sec.~\ref{sec:cfmon}), while RefinedRust
uses a refinement-type model.
Aeneas~\cite{aeneas} takes a different approach, translating a subset of Rust into a pure
functional language verified in F*, Lean, or Rocq (Table~\ref{tab:toolfull} details its coverage).
Hax~\cite{hax} also extracts THIR, but as a multi-backend translation framework targeting F*,
Rocq, Lean (via Aeneas), and domain-specific provers (ProVerif, SSProve) for security-critical
code such as cryptographic libraries; unlike the tools above, hax does not itself verify, handing
proof obligations to whichever backend is chosen---\sys{} builds directly on hax's THIR
extraction rather than competing with it.

RefinedC~\cite{SammlerLKMD021} is closest in aim to \sys{} outside Rust: Iris- and Rocq-based,
targeting systems-level C with similar architectural choices, absent Rust's ADTs and pattern
matching. VST~\cite{VST} verifies C in Rocq via separation logic; \citet{ManskyD24} re-ground it
in Iris and want an Iris-style tactic interface for entering Iris Proof Mode inside automated
proofs---a design \sys{} already supports.

\paragraph{Semantics of Rust}
RustBelt~\cite{JungKD18} gives the first foundational semantic model of $\lambda$Rust (a core
Rust-subset calculus), via a logical relation proving type-system soundness and enabling formal
reasoning about unsafe code. MiniRust~\cite{minirust}, an ongoing reference model for Rust's
memory semantics (value representation, ownership, undefined behavior), inspires \sys{}'s memory
model. Stacked Borrows~\cite{JungDKD20} and Tree Borrows~\cite{VillaniHDJ25} are candidate
operational aliasing models formalizing the provenance guarantees the borrow checker enforces,
validated dynamically via Miri and not yet adopted as a normative part of the Rust specification;
we intend to use the lifetime information they expose as an untrusted automation oracle
(Sec.~\ref{sec:memorymodel}).

KRust~\cite{krust} gives an executable K-framework semantics of a Rust subset,
automatically deriving an interpreter from rewriting rules; it provides no
program logic and no proof-assistant mechanization.

C and C++'s formal semantics predate Rust's by decades, despite sharing its concerns around
memory ownership, provenance, and undefined behavior. Cerberus~\cite{MemarianGDKRWS19} and
Krebbers'~\cite{Krebbers15} C memory model give precise accounts of these rules; we plan to draw
on their design as \sys{} develops further.

\paragraph{Effectful semantics}
\itree{}s~\cite{itrees} provide a framework for giving operational semantics
to effectful computations as denotations to coinductive interaction trees in Rocq.
\sys{} derives the soundness of its axiomatic semantics via the Program logics à la
carte framework~\cite{plalacarte};
We have completed \itree denotations and soundness proofs for a fragment of the supported
language constructs, and we are working towards covering the full semantics.

\sys{}'s control-flow sensitive specification monad~$\WPx{\cdot}$ is an instance of the
Dijkstra monad construction~\cite{d4free, d4all, maillardthesis}, which provides a
foundation for deriving WP connectives for arbitrary monadic computations.
As noted by \citet{plalacarte} Dijkstra monads do not support concurrent reasoning
and lack separation logic support for local reasoning about memory.
\sys{} addresses both gaps by constructing the underlying continuation monad atop
Iris's iProp, directly inheriting Iris's higher-order concurrent separation logic.

The two-continuation design of $\WPx{A} \triangleq (A \to \iProp) \to \mathit{CFpred} \to \iProp$
is a novel recombination of several independently well-established ideas.
Structurally, the two barrels are an instance of Thielecke's double-barrelled
CPS~\cite{thielecke01}, generalized here to a monotone predicate over the five-constructor
$\CFmonad{}$ sum rather than a single generic control/jump continuation.
Theoretically, the design falls within the Dijkstra monad lineage of~\citet{swamy13}
through~\citet{d4free} to~\citet{d4all}; \citet{d4all} explicitly reconstructs
and subsumes the classical weakest exceptional preconditions of
\citet{leinosnepscheut94} ($\mathbf{wp}(S, Q, R)$ with a normal and an exceptional
postcondition) inside Dijkstra monads.
The closest mechanized predecessor is VST's C-light continuation semantics~\cite{VST},
which carries a family of postconditions indexed by an exitkind ($\mathit{EK\_brk}$,
$\mathit{EK\_con}$, $\mathit{EK\_ret}$)---the same ``postcondition per control-flow
outcome'' idea, realized as positional tuples rather than as a single monPred-indexed
continuation threaded unchanged by bind.
The algebraic-effects reading is also applicable: $\throw$ is an algebraic operation
that commutes with $\bindCF$ (the defining property of algebraic operations per
Plotkin--Pretnar~\cite{plotkin09}), and the guard combinators at loop, function, and
match boundaries are the corresponding handlers.
The specific combination---Iris $\mathit{monPred}$-indexed control-flow barrel,
threaded unchanged by $\bindCF$, trapped by construct-specific combinators, packaged
as a single WP specification monad in Rocq---is absent from other Rocq/Iris tools
(Program Logics \`a la Carte~\cite{plalacarte}, RefinedC/RefinedRust~\cite{SammlerLKMD021,refinedrust},
Katamaran~\cite{katamaran}, Silver and Zdancewic's ITree-based Dijkstra monads~\cite{silverzd21}),
which all use a single return postcondition.

The defunctionalized continuation stacks $\scont$/$\hcont$
(Sec.~\ref{subsec:defuncont}) are an application of defunctionalization~\cite{reynolds72}
to the two-barrel WP monad; the sequential stack~$\scont$ is additionally an instance
of the reflection-without-remorse / freer-monad Kleisli sequence representation
of~\citet{kiselyov15}.

 \section{Conclusions and Future Work}\label{sec:conclusion}

We have presented \sys{}, a foundational, THIR-level verification
framework for Rust programs in the Rocq theorem prover, grounded in the
Rust Reference (\S\ref{sec:sem}) and layered atop it with a program
logic (\S\ref{sec:proglog}) and a proof automation layer (\S\ref{sec:auto}),
demonstrated on a buddy allocator case study (\S\ref{sec:casestudy}).
The layered architecture---separating axiomatic semantics, program
logic, and automation---allows language coverage and soundness to be
grown incrementally as the framework matures.

Several directions remain for future work.
The long-term goal motivating \sys{}'s design is end-to-end
verification of systems across hardware-software boundaries in a
shared Rocq foundation; this includes verified compilation of Rust to
the target hardware, for which recent work proposed a promising
path~\cite{wppresPriSC}.
We are currently scaling \sys{} to larger case studies--a unikernel
and a THIR to MIR compiler, both implemented in Rust.
At the same time, we are extending soundness of the axiomatic
semantics to all supported language constructs. On the automation side,
adapting Iris automation frameworks such as Diaframe~\cite{MulderKG22}
or Lithium~\cite{SammlerLKMD021,refinedrust} to \sys{} is a natural
next step.

\bibliographystyle{ACM-Reference-Format}

\newif\iffauxappendices
\ifdefined\PaperFullBuild
  \fauxappendicesfalse   \else
  \fauxappendicestrue    \fi

\iffauxappendices
  \appendix
\newcommand{\fauxapp}[1]{{\color{white}\section{}\label{#1}}}
  \fauxapp{sec:syn}             \fauxapp{apx:memory}          \fauxapp{apx:itree}           \fauxapp{apx:dropscopes}      \fauxapp{apx:toolcomparison}  \refstepcounter{table}\label{tab:toolfull}
  \fauxapp{apx:proglog}         \fauxapp{apx:soundness}       \fauxapp{apx:borrows}         \else
  \newpage
\appendix
  \section{Syntax}
\label{sec:syn}

The syntax of the Rust language exposed in this paper is shown in
Figure~\ref{fig:syntax-ty}\footnote{
For the sake of readability, we present a simplified-surface level syntax; the formal development is carried out on parsed AST where these syntactic concerns are not relevant.}.
For the full account, we refer the reader to the Rocq formalisation.

\paragraph{Types}
The base types of $ty$ comprise primitive types, including the boolean type \Rbool,
the character type \Rchar, and numeric types $\Rnum{sgn}{w}$,
references ($\lstmath{&} m ~ty$), and the never type (\Rnever), which
is used to formally type semantically unreachable computations.
We use the usual abbreviations \lstmath{u8}, \lstmath{i8}, etc.
Note that references are indexed by mutability ($m \in \{ \text{mutable}, \text{shared} \}$),
which makes syntactic representation more convenient to work with.

Available type formers include tuples ($\Rtuple{\overline{ty}}$), functions of arbitrary arity ($\Rfun{\overline{ty}}{ty}$),
fixed-size arrays ($\Rarray{ty}{ce}$), and slices (\Rslice{ty}).
\ff{here we need const exprs, where are these introduced? Also note that the syntax is mutually recursive}
Tuples include unit type (\Runit), which is a tuple of zero elements.
Heterogeneous product types (\lstmath{struct}) and enumerated types (\lstmath{enum})
are represented as algebraic data types (ADTs) and
indexed by their identifier ($\Radt{d}$).

\paragraph{Patterns}
Patterns are the primary means of inspecting and \emph{destructuring} values, and
appear uniformly across match expressions, let declarations, and function parameters.
Patterns include the wildcard pattern (\Rpatwild),
variable bindings ($\lstmath{ref}^?~\lstmath{mut}^?~v$) with a specified binding mode
(whether binding is by reference and at which mutability),
constant patterns, the never matching pattern (\Rnever), and
reference pattern ($\Rref{pat}$).
Note that throughout the paper we use $\text{-}^?$ to denote optional part of syntax.
Destructuring patterns over tuples ($\Rtuple{\overline{pat}}$) and ADTs ($\Radt{d}~\overline{pat}$)
decompose value into its constituent elements matched over a sequence of sub-patterns.
Finally, slice pattern ($\Rslice{\overline{pat}}$) matches an array or a slice element-wise to sub-patterns.

\paragraph{Statements}
There are two kinds of statements in Rust, expression statements and declaration statements.
\sys{} supports expression statements ($expr$) that evaluate an expression
and ignore its result, and $\lstmath{let}$ declaration statements, that introduce a set of new variables via a pattern, optionally initialized with an initializer expression.
This captures both uninitialized bindings such as \lstmath{let x;} and initialized declarations like
\lstmath{let x = 42;}, reflecting Rust's block-scoped variables.
Further, there are \emph{item} declaration statement that lexically restrict
scope of the item and impacts \emph{static} semantics of the language.
\sys{} doesn't currently support \se{Missing word.} and we conjecture these does not impact dynamic semantics.

\paragraph{Expressions} Expressions ($\Expr$) represent the core construct of the language.
The base building blocks of expressions are literal expressions ($lit$), \sys{} supports numeric literals and booleans, and path expressions that denote either local variables ($v$) or items in global scope ($gn$), e.g. constant items. There are unary operator expressions, including borrow ($\Rref e$ and $\Rrefmut~ e$), dereference ($\Rderef e$), and arithmetic ($\lstmath{-}e$) and logical ($\lstmath{!}e$) negation operators.
There are binary operator expressions ($e \odot e$) where the operators range over
binary and logical operators ($\lstmath{+},\lstmath{-},\lstmath{*},\lstmath{/},\lstmath{\%},\lstmath{|},\lstmath{^},\lstmath{<<},\lstmath{>>},\lstmath{||},\lstmath{&&}$), and
comparison operators ($\lstmath{==},\lstmath{!=},\lstmath{>}\lstmath{>=},\lstmath{<},\lstmath{<=}$).
Further, there are cast ($\Rcast{e}{ty}$) and assignment ($e~\lstmath{=}~e$) expressions.
There are array expression constructing arrays by a list of values ($\Rslice{\overline{e\lstmath{,}}}$) and by repetition ($\Rarray{e}{ce}$), and indexing expressions ($\Rslice{e}$), there are tuple ($\Rtuple{\overline{e\lstmath{,}}}$) and tuple indexing ($e\lstmath{.}n$) expression,
and there are struct expressions ($\Rstruct{gn}{\overline{gn~ (\lstmath{:} e)^?} }$) and
field access expressions ($e\lstmath{.}gn$).
There is a range of control flow expressions; a block expression
($\Rblock{\overline{stmt\lstmath{;}}~e^?}$) that sequentializes its statements and its final optional expression, a loop ($\Rloop{e}$), break ($\Rbreak$), and continue ($\Rcontinue$) expressions,
there are conditional ($\Rif~e~(\Relse~e)^?$) and pattern matching
($\Rmatch{e}{\overline{pat~\lstmath{=>}~e\lstmath{,}}})$ expressions.
Finally, there are function call ($\Rfncall{gn}{\overline{e\lstmath{,}}}$) and
return ($\Rreturn{e^?}$) expressions.

\ff{TODO: note that there is more, concerning the structure of crate/module/file - function definitions, implementations,
  and finally items}

   \section{Memory}
\label{apx:memory}

The forgetful mapping from values to runtime types is defined as follows:

\begin{definition}[$\toRTy{\text{-}}$]
{\figuresize
  \begin{align*}
\toRTy{ \valInt{s}{w} } & \triangleq~ \RTyInt{s}{w}
& \toRTy{ \valBool{b} } & \triangleq~ \RTyBool
    & \toRTy{ \valStruct{d}{fs} } & \triangleq~ \RTyStruct{d}
  \\
  \toRTy{ \valPtr{\loc} } & \triangleq~ \RTyPtr
& \toRTy{ \valTuple{\overline{r}} } & \triangleq~ \RTyTuple{\toRTy{\overline{r}}}
    & \toRTy{ \valEnum{d}{d'}{fs} } & \triangleq~ \RTyEnum{d}
\end{align*}
}
\end{definition}
   \section{ITrees and Program Logics}\label{apx:itree}
\itrees are a generic representation of effectful, possibly
non-terminating computations.
The coinductive type $\itreeER{E}{R}$ is
parametric over a type $R : \texttt{Type}$ of return values, and a
type former $E : \texttt{Type}\to\texttt{Type}$ of observable events:

\begin{definition}[ITree]
$$ \itreeER{E}{R} \ni t ::= _{\texttt{CoInductive}} \Ret{r}\;|\; \Tau{t}\;|\; \VisA{A}{e}{k}.$$
\end{definition}
\noindent
\Ret{r} represents a terminating computation that evaluates to $r:R,$
$\texttt{Tau}(t)$ represents a computation that takes a silent step and
continues with computation $t:\itreeER{E}{R},$ and
\Vis{e}{k} represents a computation that emits an event $e:
  E\;A,$ and expects an answer $a:A$ that is then passed to a
  continuation $k: A \to \itreeER{E}{R}.$
\noindent  
For a concrete example, computations that modify
heap, a partial map $\heap: \locT \rightharpoonup \mathcal{V}$
from locations to values, are represented with
$\stateE{\heap}: \texttt{Type}\to\texttt{Type}.$
\noindent
The $\stateE{\heap}$ consists of two possible events (constructors):
$\getE$ with answer type $\heap$ that returns the current state and
$\setE{s}$ with answer type \texttt{unit} that overwrites the current state with $s$.
The operation that stores the value $v$ at location $l$ of the heap,
can be represented with the following itree
\[
\Vis{\getE}{\lambda \sigma:\heap.\;
  \Vis{\setE{ \sigma [l \mapsto v]}}{\lambda u:\texttt{unit}.\;u}
      }. 
\]
Being their definition coinductive, \itrees come with a
\emph{bisimulation} relation $\approx$ that equate two itrees as long
as they differ only in (finitely many) \texttt{Tau} steps.
$\Vis{\text{-}}{\text{-}}$ induces bind that, together with return $\Ret{\text{-}}$, give rise
to a monad under equality up to finitely many $\texttt{Tau}$ steps.

\noindent
Monadic notations allow us to write the example more conveniently
as sequential code:
\begin{flalign*}
  &\store{l}{v} := \\
  &\qquad \sigma \gets \trigger{\getE}; \\
  &\qquad u \gets \trigger{\setE{\sigma [l \mapsto v]}}; \\
  &\qquad \Ret{u}
\end{flalign*}
with $\trigger{E} := \Vis{E}{\lambda x.\;\Ret{x}}.$

\subsection*{Computational and logical interpretation of \itrees}
The events $E$ in \itreeER{E}{R} can be given a computational
interpretation by handling them in a monad $M.$ A \emph{computational}
handler\footnote{Usually just called handler, however we want to
distinguish them from the logical handlers introduced
by~\citet{plalacarte}.} $H: \forall A.\; E\;A \to M\;A$ can then be
recursively extended over $t : \itreeER{E}{R},$
thus interpreting \itrees into monadic computations.
The handler for $\texttt{stateE}_{\heap}$ that maps $\getE,
\setE{\cdot}$ to the get and set operation of a state monad can be
used to interpret $\store{}{}$ as a stateful computation.

\medskip
\noindent
The events $E$ in \itreeER{E}{R} can be given also a logical
interpretation by specifying pre- and post-conditions for the
triggering of the event.
For example, the pre-condition for triggering~$\setE{\sigma [l \mapsto
    v]}$ is to have ownership of the old state $\sigma,$ while its
post-condition -- for every answer $u : \texttt{unit}$ -- is to give
back the ownership of the new state. Similarly, the pre-condition to
$\getE$ is ownership of the state but the post-condition
requires the state to be given back unchanged.

\noindent
Pre- and post-conditions to the triggering of an event can be given in
a weakest pre-condition style by means of a \emph{logical} handler
$H_A(e, \Psi) := \mathit{Pre} \ast (\forall a : A.\; \mathit{Post}(a) \wand \Psi(a)).$
From a logical handler $H$ it's then possible to define a weakest
pre-condition calculus for itrees~\cite{plalacarte}. For an itree $t$
and a (resource-aware) predicate $\Phi$ the weakest pre-condition
$\wpi{H}{t}{\Phi}$ is defined by~\footnote{Everything is wrapped in
the update modality for updating Iris ghost state when
necessary\cite{iris}.} 

\begin{equation*}
  \wpi{H}{t}{\Phi} := \updateMod
\begin{cases}
    \Phi(r)                                   & \text{if}~ t = \Ret{r}  \\
    \wpi{H}{t'}{\Phi}                         & \text{if}~ t = \Tau{t'}  \\
    H_A(e, \lambda\;a.\;\wpi{H}{k(a)}{\Phi})  & \text{if}~ t = \VisA{A}{e}{k}
  \end{cases}
\end{equation*}
If $t=\Ret{r}$ then the returned value $r$ must satisfy $\Phi,$ while
if $t=\Tau{t'}$ the same weakest pre-condition must be satisfied by
$t'.$ For $t=\VisA{A}{e}{k}$ it's necessary to satisfy the
pre-conditions for triggering $e$ -- specified by the handler $H$ --
and to show that the post-condition -- specified by $H$ as well --
implies $\wpi{H}{k(a)}{\Phi}$ along every possible continuation
$k(a)$.

Recalling that
$\trigger{\getE} := \Vis{\getE}{\lambda \sigma.\Ret{\sigma}},$
the readers can convince themselves that a proof of
$\wpi{H}{\store{l}{v}}{\Phi}$ consists of the following steps:
\begin{enumerate}
  \item prove ownership of the current state $\sigma$ (pre-condition
    for \getE) and get as post-condition for the continuation $ u
    \gets \trigger{\setE{\sigma [l \mapsto v]}}; \Ret{u}$ the state
    unchanged and the ownership of the state
  \item prove again ownership of $\sigma$ (pre-condition for
    \setE{\cdot}) get as post-condition for $\Ret{u}$ that $\Phi$
    must hold in the updated state $\sigma [l \mapsto v].$
\end{enumerate}
From steps (1) and (2) above we immediately derive the following rule
\begin{equation*}\label{eq:wp-store}
\inferrule[wp-store]
    { (l \mapsto v) \wand\; \Phi }
    { \wpi{H}{\store{l}{v}}{\Phi}  }
\end{equation*}

\subsection*{Modularly Extending Logics for \itrees}\label{subsec:plalacarte} 
The behavior of Rust programs can sometimes be
undefined~\cite{RustLang}.
We therefore extend $\stateE{\heap}$ -- i.e. \getE\; and \setE{\cdot}
-- with an event modeling an undefined behavior, \ubE\; and extend
$\store{}{}$ to trigger \ubE\; on invalid locations,
\begin{flalign*}
  &\storeorub{l}{v} := \\
  &\qquad\texttt{match}\; \mathit{valid}(l) \;\texttt{with} \\
  &\qquad\texttt{ | True  } \Rightarrow \store{l}{v} \\
  &\qquad\texttt{ | False } \Rightarrow \trigger{\texttt{ubE}} \\
  &\qquad\texttt{end.}
\end{flalign*}
More rigorously, while
$\store{l}{v}:\itreeER{\stateE{\heap}}{\mathcal{V}},$ its extension
with \ubE\; has type
\[
   \storeorub{l}{v}:\itreeER{(\stateE{\heap}+'\ubE)}{\mathcal{V}}
\]
where $+'$ is the \emph{sum} of two event types.

\noindent
Since \ubE\; models an undefined behavior its logical handler
$H_{ub}$ imposes to prove false in the post-condition.
We can combine $H_{ub}$ with $H$ -- logical handler for \stateE{\heap}
-- with the operation $\oplus$ and relying on an adequacy
lemma~\cite{plalacarte}.
Finally, we can extend the rule $\textsc{wp-store}$ to
\begin{equation*}\label{eq:wp-store-ub}
\inferrule[wp-store-ub]
    { \ulcorner \mathit{valid}(l) \urcorner \ast ~(l \mapsto v) \wand\; \Phi }
    { \wpi{H\oplus H_{ub}}{\storeorub{l}{v}}{\Phi}  }
\end{equation*}
\subsection*{Program Logics from \itrees semantics}
So far we recalled how to define a weakest-precondtion calculus for
terms of type \itreeER{E}{R}, for certain events $E$ and return values
$R$. We also recalled how to modularly extend the calculus when
extending $E$ with further events -- \ubE\; in our example above.
Not surprisingly, the wp-calculus built for \itreeER{E}{R} is also a
wp-calculus for the expressions of a language denoted with
\itreeER{E}{R}.
As an example, we can give $\lstmath{let mut x = e}$ a denotation
\itreeof{\cdot} into \itrees with events $\stateE{\heap}+'\ubE$ and
return values $\mathcal{V}$ from~\Cref{fig:valrtyloc},
\begin{flalign*}
  \itreeof{\lstmath{let mut x = e}} :=\; 
  & l_x \gets \itreeof{\lstmath{x}}; \\
  & v_e \gets \itreeof{\lstmath{e}}; \\
  & \storeorub{l_x}{v_e}
\end{flalign*}
By interpreting $\mathrm{wp}\; e\;\xi$ as
$\wpi{H\oplus H_{ub}}{\itreeof{e}}{\xi}$ we can show~\todo{double-check
  and prove it}
\begin{equation*}\label{eq:wp-example}
  \inferrule[Let-lemma]
    { x \mapsto l\;\ast \quad e \mapsto v\;\ast \quad (l \mapsto v) \wand\; \Phi }
    { \mathrm{wp}\; \lstmath{let mut x = e}\;\xi }
\end{equation*}
by simply unfolding the denotation $\itreeof{\lstmath{let mut x = e}}$
and applying $\textsc{wp-store-ub}.$

\subsection{ITrees Semantics for THIR and Axioms Validated}\label{subsec:axioms2lemmas}
We are currently establishing the soundness of each of the axioms
from~\cref{sec:sem} by modularly extending the \itrees semantics to
all terms in $\Expr$ and $\Stmt$.
The type of returned values is $\mathcal{V}$ from~\Cref{fig:valrtyloc}
and the event type we started with is
\[ 
   \thirE := \stateE{\heap}\;+'\ubE\;+'\stepE
\]
where $\stateE{\heap}$ models stateful computation over a heap as
described above, \ubE\; models an undefined behavior and \stepE\
consists of a single event with answer type $(),$ with no
computational effect but that, on the logic side, allows for \lobind\
induction~\cite[Section 6.1]{plalacarte}.

\noindent
No concrete implementation of the memory model is provided. The heap
is modeled as a partial function form locations to values $\heap :
\locT \rightharpoonup \mathcal{V}$ and $\locT$ as well as the
\emph{points-to} predicates are not implemented but coincides with the
axiomatization from~\Cref{sec:memorymodel}.

\noindent 
We define
\[
   \itreeof{e}: \itreeER{\thirE}{\mathcal{V}}
\]
for every THIR \texttt{stmt} $e.$
We then give a logical interpretation to $\itreeof{e}$ by means of the
logical handler $H_{\texttt{THIR}}$ obtained combining the handlers
for \stateE{\heap},\,\ubE\; -- described above -- and the handler for
\stepE\; that just wraps a predicate $\Phi$ under $\vartriangleright,$
the \emph{``next''} modality.

\smallskip
\noindent
Notice that for a THIR expression $e\in\Expr$,
$\wpi{H_{\texttt{THIR}}}{\itreeof{e}}{\Phi} : (\mathcal{V} -> iProp)
-> iProp,$ so that in order to recover $\WPvexpr$
from~\Cref{def:corepreds} we just need to apply the exception monad
transformer $\wpi{H_{\texttt{THIR}}}{\itreeof{e}}{\Phi}.$
\crmn{ I don't see this as a limitation in principle as all the spec are given for $\Phi: A -> iProp$ and not $\CFmonad{A} -> iProp$}
We instantiate \WPpexpr, \WPcexpr, \WPstmt from~\Cref{def:corepreds}
in a similar way.

\noindent
Several rules from~\Cref{sec:sem} such as $\textsc{wp-stmt-expr}$
becomes immediate consequence of \itrees providing a denotational
semantics: the denotation of a value (or place) expression is just the
denotation of it's embedding into \texttt{stmt}.

\noindent
Despite the simplicity of the model, we were able to instantiate the
rules from~\Cref{sec:sem} and several of the rules
from~\Cref{subsec:direval}.
   \section{Drop Scopes}\label{apx:dropscopes}

The Rust Reference~\Rhrefdoc{destructors.html\#r-destructors.scope}{[RR §6.1--6.5]}
defines five kinds of \emph{drop scopes} that determine when owned values are
destroyed: the function scope (outermost), block scopes, match arm scopes,
statement scopes, and expression (scrutinee) scopes.
When a scope exits---by normal completion, \lstmath{return}, \lstmath{break}, or
\lstmath{continue}---every value owned by that scope is dropped in reverse declaration order.
Scopes are nested; early exits must drain all scopes from the innermost outwards.

We give the function frame~$\Xi$ (Sec.~\ref{sec:sem}) the structure needed to support
this: rather than a single flat map, $\Xi$ is a \emph{stack} of scopes, each holding
the name-to-location bindings introduced at that nesting level. Threading $\Xi$
through the WP calculus this way gives correct lexical shadowing (an inner scope's
bindings are found before an outer scope's) and, as the same stack structure is
already exactly the information needed to know what to clean up and in what order,
supports drop-scope cleanup with no separate mechanism.

\begin{definition}[Function frame, $\Xi$]\label{def:scopestack}
  \figuresize
  \[
    \Xi \ni \xi ::= \nil
    \mid \ScopeNamed~\ell^{?}~\overline{(i,\loc)} :: \xi
    \mid \ScopeTmp~\overline{\loc} :: \xi
  \]
  where $i$ ranges over local variable names, $\loc : \Loc$, and $\ell^{?}$ is an
  optional break/continue label.
\end{definition}

\noindent
$\ScopeNamed$ entries are pushed by block, arm, and loop scopes and carry the
name-location bindings of the locals owned by that scope; $\ScopeTmp$ entries are
pushed by statement scopes and match expressions to track (unnamed) temporaries.
A \emph{single} interleaved list preserves the exact nesting order, enabling an
early exit to drain both named and temporary scopes in one downward pass.
The distinction is operationally relevant: \lstmath{break}/\lstmath{continue}
target $\ScopeNamed$ entries by depth, while $\ScopeTmp$ entries are always
drained as collateral when passing through them. Variable lookup, $\xi[id] = \ell$
(cf. \WPPexprVar, Fig.~\ref{fig:wp-pexpr}), searches $\xi$ top-down for the first
$\ScopeNamed$ entry binding $id$.

\begin{figure}\RhrefToDocstrue
  \figuresize
  \begin{flushleft}
    \boxed{$$\WPblock : \text{list}~\Stmt \to \Expr^? \to \Xi \to \WPx{\Val}$$}
  \end{flushleft}
  \begin{align*}
    \WPblock~(s \cons ss)~e^?~\xi &\triangleq
      \xi' \leftarrow \WPstmt~s~(\ScopeNamed~\nil :: \xi)~;~
      \WPblock~ss~e^?~\xi' \\
    \WPblock~\nil~e~\xi &\triangleq \WPvexpr~e~\xi \\
    \WPblock~\nil~\emptyset~\xi &\triangleq \retCF~\valUnit
  \end{align*}
  \begin{flushleft}
    \boxed{$$\WPdrop : \Loc \to \Type \to \Xi \to \WPx{\Xi}$$}
  \end{flushleft}
  \begin{mathpar}
    \mprset{flushleft}
    \inferrule[\WPDropInit]
              { \loc \pointsto v \ast \WPcallDrop~\mathit{ty}~v~\xi~Q }
              { \WPdrop~\loc~\mathit{ty}~\xi~Q }

    \inferrule[\WPDropUninit]
              { \loc \pointstoUninit~\mathit{ty} \ast Q~\xi }
              { \WPdrop~\loc~\mathit{ty}~\xi~Q }
  \end{mathpar}
  \caption{Updated $\WPblock$ and $\WPdrop$ with the stack-structured frame~\Rhrefdoc{destructors.html\#r-destructors.scope}{[RR §6.1--6.5]}}
  \label{fig:wp-dropscopes}
\end{figure}\RhrefToDocsfalse

\paragraph{Block scopes}
The updated $\WPblock$ (Fig.~\ref{fig:wp-dropscopes}) pushes a fresh $\ScopeNamed$
entry before evaluating each statement.
Both $Q_\text{seq}$ and $Q_\text{rt}$ are implicitly wrapped to drain the scope at exit:
on normal completion, the top $\ScopeNamed$ entry is popped and the locations of its
bindings are dropped in reverse order; on any early exit ($\BreakN{n}$, $\ContinueN{n}$, $\Return$),
the scope is drained before the exit is rethrown to the enclosing handler.
$\ScopeTmp$ entries are pushed and popped by $\WPstmt$ around each statement,
dropping temporaries created during evaluation before proceeding.

The denotation of each \lstmath{let}-binding emits a $\StorageLive$ event, producing
a fresh location $\loc$, and the denotation appends the pair $(i,\loc)$ to the top
$\ScopeNamed$ entry---making cleanup fully explicit in the ITree with no implicit
scope state in the handler.

\paragraph{Match arm scopes}
A match arm scope owns pattern bindings and guard bindings for a single arm.
We split the existing $\WParm$ into $\WPguard$ and $\WParm$:
$\WPguard$ pushes a fresh $\ScopeNamed$ entry and handles all failure paths---if the
pattern does not match or a guard fails, the scope (with any partially introduced
bindings) is drained before the next arm is attempted;
$\WParm$ receives an already-matched, already-guarded stack and runs the arm body,
popping the $\ScopeNamed$ entry pushed by $\WPguard$ after completion.
The match expression itself pushes a $\ScopeTmp$ entry for scrutinee temporaries,
which outlive individual arms~\Rhrefdoc{destructors.html\#r-destructors.temporary-lifetime}{[RR §6.4]}.

\paragraph{Loops}
$\WPvexpr$ for \lstmath{loop} pushes a $\ScopeNamed$ entry and uses
the one-unfolding semantics from Sec.~\ref{sec:expr}: the sequential continuation of
the unfolded body recurses on the loop (next iteration), and $Q_\text{rt}$ is wrapped
to intercept $\BreakN{n}$ targeting this loop (depth $n=1$)---draining the loop's
scope and continuing sequentially---or, for $n>1$, to decrement the depth and rethrow
to the outer handler.
Nested loops are thus handled without extending the $\CFmonad{}$ type: the
$Q_\text{rt}$ continuation bakes in the dispatch by depth alone.

\paragraph{Drop semantics}
$\WPdrop$ (Fig.~\ref{fig:wp-dropscopes}) dispatches on whether the location is
initialized or not, avoiding runtime drop flags.
For initialized locations (\WPDropInit), $\WPcallDrop$ runs the destructor: for
types implementing \lstmath{Drop} this calls the user-defined method and then
recursively drops fields; for other compound types it directly drops fields.
For types implementing \lstmath{Drop}, this requires an assumed specification
of the type's \lstmath{drop} method to be available and dischargeable, consuming
the value's ownership and releasing the specification's postcondition as a
resource; for compound types without a \lstmath{Drop} implementation, it instead
requires the ownership of the value's fields to be directly available for
recursive dropping. Determining which of these two cases applies to a given
type---and, in the former case, locating the specification to instantiate---is
static, type-directed information already available in THIR; threading it
through the WP calculus, analogous to trait-method resolution
(Sec.~\ref{sec:relatedW}), requires a typing environment that we leave out of
scope for this paper.
For uninitialized locations (\WPDropUninit), no destructor call is needed; the
axiom immediately hands off to $Q$.
The $\loc \pointsto v$ vs.\ $\loc \pointstoUninit$ distinction is the
separation-logic counterpart of MIR's runtime drop flags, handled purely at the
proof level.
$\WPpat$ (Fig.~\ref{fig:wp-pat}) already produces $\loc \pointstoUninit$ for
moves, so the correct branch is always determined by the available separation-logic
resources.

\paragraph{Function scope}
The function scope is handled by extending $\WPfn$ with an outermost $\ScopeNamed$
entry for function parameters, which are treated uniformly as block-scoped locals.
This makes function-scope cleanup uniform with block cleanup and requires no
separate mechanism.

\paragraph{Drop scope events}
Drop scopes introduce three further events to the ITree vocabulary (Sec.~\ref{sec:itreemodel}):
\begin{itemize}
  \item $\StorageLive : \mathit{DefId} \to \mathit{size} \to \mathit{align} \to \Loc$ ---
        allocates a stack slot and returns a fresh location; its logical handler
        produces $\loc \pointstoUninit$ in the Iris context.
  \item $\StorageDead : \Loc \to \mathtt{unit}$ ---
        marks a stack slot dead; its logical handler consumes the
        $\loc \pointsto v$ or $\loc \pointstoUninit$ predicate.
  \item $\DropEvent : \Loc \to \Type \to \mathtt{unit}$ ---
        runs the destructor of the value at the given location.
        For types implementing \lstmath{Drop}, this calls the user-defined
        destructor and then recursively drops fields; for other compound types
        it directly drops fields.
        The compiler intrinsic \lstmath{ptr::drop_in_place<T>} is given a
        derived specification: it resolves its pointer argument to a location
        $\loc$ and emits $\DropEvent(\loc, T)$, semantically identical to the MIR
        \lstmath{drop} terminator.
\end{itemize}

The modular extension approach ensures that the logical handlers for these events
are adequate with respect to the existing handlers for \texttt{heapE} and \ubE:
the separation-logic reasoning principles for $\StorageLive$/$\StorageDead$
correspond directly to the fractional ownership model (Sec.~\ref{sec:memorymodel}),
and $\DropEvent$ expands into further $\StorageLive$/$\StorageDead$/\texttt{heapE}
events whose handlers are already justified.
A $\StorageLive$/$\StorageDead$-pair per loop iteration ensures that pointers into
a previous iteration's stack slot become invalid, which is the correct provenance
behavior for stack allocations.
   \section{Rust Verification Tool Comparison}\label{apx:toolcomparison}

Table~\ref{tab:toolfull} expands Table~\ref{tab:toolcompact} (Sec.~\ref{sec:relatedW})
with the theory, prover/solver backend, and language coverage of each tool.

\begin{landscape}
\setlength{\tabcolsep}{0.5em}
\begin{table}[h]
\footnotesize
\centering
\begin{tabular}{@{}p{0.12\linewidth}p{0.12\linewidth}p{0.06\linewidth}p{0.24\linewidth}p{0.15\linewidth}p{0.19\linewidth}@{}}
\toprule
\textbf{Tool} & \textbf{Level} & \textbf{Found.} & \textbf{Theory / backend} & \textbf{Automation} & \textbf{Coverage} \\
\midrule
Miri~\cite{miri} & MIR, dynamic execution & --- &
  Operational (Rust Abstract Machine; Stacked/Tree Borrows) &
  Automatic (concrete execution) &
  unsafe code, most of \texttt{std}; single execution path only \\
Kani~\cite{kani,kani2026} & MIR $\to$ Goto-C &
  No &
  Bounded model checking (CBMC); SAT (MiniSat) or SMT (Z3, cvc5, bitwuzla) &
  Automatic &
  unsafe code, panics, overflow; no concurrency \\
Verus~\cite{verus} & Rust (ghost code) $\to$ SMT-LIB &
  No$^\dagger$ &
  Linear ghost types / modes, FOL; Z3 (via SMT-LIB) &
  Automatic (SMT) &
  unsafe code (via ghost permissions), concurrency \\
Prusti~\cite{prusti} & MIR $\to$ Viper~\cite{viper} &
  No &
  Implicit dynamic frames; Silicon (symb.\ exec., Z3) or Carbon (VCGen, Boogie) &
  Mostly automatic (SMT); loop invariants sometimes manual &
  primarily safe Rust; no concurrency \\
RustHorn\-(Belt)~\cite{rusthornbelt} & Rust subset $\to$ Horn clauses &
  No$^\dagger$ &
  CHC solving; prophecies (RustHornBelt: parametric prophecies via Iris) &
  Automatic (CHC solver); RustHornBelt needs a front-end verifier (e.g.\ Creusot) &
  safe APIs over unsafe internals (\texttt{Vec}, iterators); concurrency (\texttt{Mutex}, spawn/join) \\
Creusot~\cite{creusot,coma25} & MIR $\to$ Coma (Why3) &
  No &
  Deductive VCGen, prophecies for mutable borrows; Why3 $\to$ Alt-Ergo/CVC/Z3 &
  Mostly automatic (SMT); invariants sometimes manual &
  generics, trait bounds; primarily safe Rust \\
Flux~\cite{flux} & MIR, rustc plugin &
  No &
  Refinement (liquid) types; liquid-fixpoint (Z3 internally) &
  Highly automatic (liquid inference) &
  safe Rust, ownership-aware strong updates \\
Refined\-Rust~\cite{refinedrust} & MIR &
  Yes &
  Refinement types with an Iris semantic model in Rocq; Lithium proof search &
  High (Lithium automation); manual Iris proof for complex cases &
  safe and unsafe Rust, generics; single-threaded \\
Aeneas~\cite{aeneas} & Rust subset $\to$ pure functional &
  Semi$^\ast$ &
  Functional translation of ownership/borrowing; F*, Lean, or Rocq &
  Automatic translation; manual/interactive proof &
  excludes unsafe code, concurrency; loop/trait coverage has expanded since ICFP'22$^\ddagger$ \\
\textbf{\sys{}} & \textbf{THIR} & \textbf{Yes} &
  \textbf{Hoare-type theory + WP specification monad, atop Iris; Rocq} &
  \textbf{Syntax-driven (proof-search tactics)} &
  \textbf{safe and unsafe (uniform), control flow, drop; traits partial, concurrency inherited from Iris but not yet exercised} \\
\bottomrule
\end{tabular}
\caption{Full comparison of Rust verification tools against \sys{}.
$\dagger$~Verus's core linearity/mode metatheory is proved for a small calculus in the
paper, but each verification instance is still discharged by trusting Z3/cvc5's
result, so per-program verification is not foundational in \sys{}'s sense.
RustHornBelt mechanizes, in Rocq and Iris, the soundness of the Horn-clause
encoding with respect to $\lambda$Rust semantics---a genuine foundational
metatheorem---but each concrete verification run still trusts an external
solver's answer (a CHC solver directly, or an SMT solver via a front-end
verifier such as Creusot) rather than producing a kernel-checked certificate.
$\ddagger$~As of the cited ICFP'22 publication, Aeneas's symbolic semantics did not support
loops or most trait usage; loop support has since been substantially extended in Aeneas's
public development (only a narrow nested-loop corner case remains unsupported), while unsafe
code and concurrency remain excluded.
$\ast$~Aeneas's F*/Lean/Rocq proofs about the functional translation are
kernel-checked, but a soundness argument connecting the translation to Rust's actual
semantics is, per the authors, still future work---not yet established in any form,
mechanized or otherwise.
}
\label{tab:toolfull}
\end{table}
\end{landscape}
   \section{Program Logic and Automation Details}\label{apx:proglog}

This appendix gives the full versions of the program logic and automation
material summarized in Secs.~\ref{sec:proglog} and~\ref{sec:auto}.

\subsection{Sealing: Full Set of Block Lemmata}

In the case of $\WPblockSealed$, we provide three lemmata for controlled
unfolding (Fig.~\ref{fig:proglog-full}): \WPBlockCons{} for blocks with a
leading statement, and \WPBlockNilExpr{} and \WPBlockNilNoExpr{} for empty
blocks, depending on whether a trailing expression is present. The first rule
transforms a block with a leading statement into the weakest precondition of
that statement, sequentially followed by the (sealed) WP of the remaining
block, enabling evaluation of the leading statement first while keeping the
remaining block opaque until needed. \WPBlockNilExpr{} and
\WPBlockNilNoExpr{} handle the base case, dispatching on whether a trailing
expression is present. This technique yields a cleaner and more readable
proof state and significantly reduces proof term size and checking time.

\begin{figure}[h]
  \figuresize
  \newbox\bBlockNilE
  \setbox\bBlockNilE\hbox{\Rblock{e~}}
  \newbox\bBlockNilNE
  \setbox\bBlockNilNE\hbox{\Rblock{}}
  \newbox\bUnit
  \setbox\bUnit\hbox{\Runit{}}
  \begin{mathpar}
    \mprset{flushleft}
    \inferrule[\WPBlockNilExpr]
              {\quad\WPvexpr~e~\xi }
              {\WPstmt~\usebox\bBlockNilE{}~\xi}

    \inferrule[\WPBlockNilNoExpr]
              {\quad\retCF{\usebox\bUnit{}} }
              {\WPstmt~\usebox\bBlockNilNE{}~\xi}

    \inferrule[\WPPatConst]
              { c \leftarrow \WPcexpr~ce ~;
                \\\\
                \delayedCpat ~ c ~ e ~ \xi
              }
              {\WPpat~ce~e~\xi}
  \end{mathpar}
  \caption{Remaining program logic lemmata, complementing Fig.~\ref{fig:proglog}}
  \label{fig:proglog-full}
\end{figure}

\subsection{Reordering Evaluation: Full Derivation}

In the semantics, WP accumulation proceeds by structural traversal of the AST,
which does not always align with the order in which a programmer
would naturally reason about evaluation.
Consider matching a value against a literal pattern. In the core semantics,
pattern matching first accumulates the WP of the scrutinee expression via $\WPvexpr$,
and only afterward accumulates the weakest precondition of the pattern via
$\WPpat$.
However, the literal appearing in a pattern is itself a weakest precondition of
a constant expression, $\WPcexpr$, and contrary to the programmer's intuition
this constant has not yet been evaluated when the scrutinee is processed.
In idiomatic reasoning, one would expect the constant to be evaluated first,
and only then proceed to evaluate the scrutinee and perform the match.

To support this reasoning pattern, we introduce a derived rule for literal patterns that reverses
the apparent order of evaluation. Technically, we introduce a sealed definition
$\delayedCpat$ that encapsulates the continuation corresponding
to the subsequent evaluation of the scrutinee and
a derived inference rule for literal patterns (\WPPatConst, Fig.~\ref{fig:proglog-full}):

\begin{definition}[Reordering constant patterns]
  \begin{align*}
    \delayedCpat \triangleq
    \text{if } cv = v \text{ then } \retCF{\xi}
                  \text{ else } \PatNoMatch
  \end{align*}
\end{definition}

This rule accumulates the WP of the constant expression $ce$ of the pattern first,
binding its value $c$ into a continuation,
and subsequently accumulating the WP of the scrutinee expression $e$.
The pattern is checked once both values are available.
Crucially, this reordering is sound because the evaluation of the constant expression
is independent of the scrutinee--there are no side effects or dependencies that would constrain the order.

\subsection{Join-Point Reasoning: The General Case}

Join-point reasoning is not limited to if expressions. In the general case,
where evaluation dispatches over an arbitrary scrutinee, the same style of
reasoning applies to \lstmath{match} expressions: each arm is discharged
independently against a shared intermediate postcondition $R$, and the
continuation following the match is proved once against $R$ composed with
the frame not consumed by whichever arm was taken, exactly as for
\WPIfJoin{} (Fig.~\ref{fig:proglog}) but with $n$-ary case analysis in place
of the binary \lstmath{if}/\lstmath{else} split.

Reasoning about \lstmath{loop} expressions also requires join-point
reasoning, though the join point here connects loop \emph{iterations} rather
than branches: because loops are modeled by one-unfolding
(Sec.~\ref{sec:sem}), a user-supplied loop invariant serves simultaneously as
the postcondition establishing the unfolded body's obligations and as the
precondition available at the start of the next iteration---the same
common-precondition/shared-postcondition pattern as \WPIfJoin{}, instantiated
with the invariant playing the role of $R$ and the loop's continuation (after
a $\Break$) playing the role of the final postcondition $Q$.

\subsection{Proofmode Tactics on Continuation Stacks}

The proofmode tactics (Sec.~\ref{sec:auto}) operate directly on the
defunctionalized continuation stacks (Sec.~\ref{subsec:defuncont}).
A $\bindCF$ application pushes a new $\SKbind$ frame, and a $\retCF$ application pops
the top frame (or resolves directly at the base case), both via $\seqK\text{-unfold}$,
instead of building $\lambda a.\,f\,a\,Q_\text{seq}\,Q_\text{rt}$ closures.
$\mathit{tac\_throw}$ dispatches on the head $\mathit{HK}_X$ constructor of $\hcont$
via $\rtK\text{-unfold}$, firing or popping the top handler on the matching control-flow tag.
Scope installers $\mathit{tac\_guard}_{X\text{-CF}}$ push one $\mathit{HK}_X$ handler
and one $\SKbind$ frame.
The dispatch table of the main automation loop is structurally unchanged; only the
continuation arguments in tactic right-hand sides become $\seqK\,(\rtK\,H)\,S$ / $\rtK\,H$.
   \section{Soundness Status and Validation}\label{apx:soundness}

This appendix itemizes the soundness status of the axiomatic semantics
(Sec.~\ref{sec:sem}) against the \itrees denotation (Sec.~\ref{sec:itreemodel},
Appendix~\ref{apx:itree}), and describes how the axioms are validated against
the intended semantics of Rust independently of that proof.

\subsection{Per-Axiom Adequacy Status}

Table~\ref{tab:adequacy} lists every named rule in Figs.~\ref{fig:wp-pat}--\ref{fig:wp-vexpr2}
and its adequacy status against the \itrees denotation. As established in
Sec.~\ref{sec:itreemodel}, completed proofs are final: the compositional
construction means that extending coverage never requires revisiting an
already-adequate rule, only proving adequacy for the newly introduced event
types. \emph{Completed} entries have a closed Rocq proof relating the axiom to
its \itrees denotation, following the pattern illustrated for
\textsc{Binop-add} and \textsc{Let-addV-lemma} in
Sec.~\ref{sec:itreemodel}; \emph{Pending} entries are stated as axioms only.

\begin{table}[h]
\figuresize
\centering
\begin{tabular}{ll}
\toprule
\textbf{Rule} & \textbf{Status} \\
\midrule
\WPVexprLiteralBool, \WPVexprLiteralInt & Completed \\
\WPVexprUnOpNeg, \WPVexprBinOp & Completed \\
\WPStmtExpr & Completed \\
\WPStmtLet & Completed \\
\WPVexprVar, \WPPexprVar & Pending \\
\WPVexprBorrowPlace, \WPVexprBorrowValue, \WPPexprBorrow & Pending \\
\WPVexprTuple, \WPVexprAdt, \WPPexprTupleField, \WPPexprAdtField & Pending \\
\WPVexprIf, \WPVexprBlock & Pending \\
\WPVexprMatchV, \WPVexprMatchP & Pending \\
\WPVexprLoop, \WPVexprBreak, \WPVexprContinue & Pending \\
\WPVexprCallGlobname, \WPVexprReturn & Pending \\
$\WPpat$ (Fig.~\ref{fig:wp-pat}, all clauses) & Pending \\
\bottomrule
\end{tabular}
\caption{Adequacy status of the axiomatic semantics against the ITrees
denotation. Completed proofs cover the pure fragment (literals, unary and
binary operators) and heap-manipulating statements (Sec.~\ref{sec:itreemodel}).}
\label{tab:adequacy}
\end{table}

\subsection{Test-Suite Construct Coverage}

Table~\ref{tab:eval} (Sec.~\ref{sec:auto}) reports the ten collections of the
synthetic test suite by source and proof size; each collection's name
indicates the constructs it targets---e.g. \texttt{arith} and \texttt{unary}
exercise \WPVexprBinOp/\WPVexprUnOpNeg, \texttt{cf} exercises
\WPVexprLoop/\WPVexprBreak/\WPVexprContinue, \texttt{ifs} exercises
\WPVexprIf join-point reasoning, \texttt{matching} exercises $\WPpat$ and
\WPVexprMatchV/\WPVexprMatchP, \texttt{tuple} exercises
\WPVexprTuple/\WPVexprAdt and the corresponding place-expression rules, and
\texttt{impls} exercises \WPVexprCallGlobname. The full test suite contains
further collections beyond these ten, exercising language features (e.g.
generics, trait dispatch) that the main body omits for space reasons
(Sec.~\ref{sec:relatedW}); those collections are not reflected in
Table~\ref{tab:eval} or in this appendix's per-construct breakdown.
This test suite validates the
axioms against the intended semantics of Rust independently of the adequacy
proofs in Table~\ref{tab:adequacy}: adequacy establishes that an axiom is
consistent with \emph{some} operational model, while the test suite checks
that the axiom captures Rust's \emph{actual} behavior as documented in the
Reference and the standard library.

\subsection{The Test-Suite Oracle}

Each test in the suite is a Rust source file annotated with a hand-authored
functional specification that states the outcome a programmer familiar with
the Rust Reference and standard library documentation would expect (e.g. the
postcondition of \Rfn{example} in Fig.~\ref{fig:motivation-function}, or the
well-formedness invariant of the buddy allocator in
Sec.~\ref{sec:casestudy}). The oracle is this hand-authored specification, not
an independent execution of the program: a test passes when the axiomatic
semantics lets us discharge the specification via \sys{}'s automation or
manual proof. A discharged proof is therefore evidence that the axioms agree
with the documented, human-understood semantics of the construct under test,
complementing---but not replacing---the \itrees adequacy proofs of
Table~\ref{tab:adequacy}, which check agreement with a specific operational
model rather than with the Reference's prose.

\subsection{Extraction Roadmap}

The \itrees library supports extracting an \itreeER{E}{R} denotation to
executable OCaml when the event handlers for $E$ are given a concrete
interpretation~\cite{itrees}; Vellvm's LLVM IR semantics is extracted this
way to obtain an executable interpreter validated against the reference
LLVM toolchain~\cite{vellvm2021}. We intend to follow the same route for
\sys{}'s \itrees denotation: extracting an interpreter for the covered THIR
fragment would let us differentially test axiom coverage against
\texttt{rustc}'s and Miri's own execution. Concretely, we plan to annotate
each test in the suite with its expected input/output pairs and run these as
ordinary Rust \texttt{\#[test]} functions over the crafted inputs already
used to state the specifications described above, alongside the manual
oracle. This would upgrade the test suite's validation from a hand-authored
oracle to an automated, executable one, without requiring changes to the
axioms or their adequacy proofs.
   \section{Borrows, Lifetimes, and the Trusted Computing Base}\label{apx:borrows}

This appendix works a mutable-reborrow example through the pattern rules of
Sec.~\ref{sec:pat} (Sec.~\ref{sec:memorymodel} states the general principle),
itemizes \sys{}'s trusted computing base (TCB), and sketches how the buddy
allocator case study (Sec.~\ref{sec:casestudy}) threads \lstmath{\&mut self}
ownership across its call graph.

\subsection{Worked Example: A Mutable Reborrow}

Consider binding a mutable reference to an existing local, e.g.\ matching
the pattern \lstmath{ref mut r} against a place holding value $v$ at some
location. The relevant clause of $\WPpat$ (Fig.~\ref{fig:wp-pat}) is
\[
  \WPpat~(\lstmath{ref}~\lstmath{mut}~i)~v~\xi ~\triangleq~
  \exists \loc.\ \loc \pointsto v \ast
  \big(\loc \pointsto v \mwand
    \forall \loctmp.\ \loctmp \pointsto \lstmath{ptr}~\loc \mwand \retCF{(\xi \cup \framelet{i}{\loctmp})}\big).
\]
Reading this as a borrow: the rule demands full ownership $\loc \pointsto v$
of the location being borrowed---not a fraction---reflecting that a mutable
borrow is exclusive. It immediately returns that same fact behind a magic
wand, so the continuation obtains it back unchanged; what is consumed and not
returned is only the ability of the \emph{current} proof step to use
$\loc \pointsto v$ directly; the new binding $i$ instead points to a fresh
location $\loctmp$ holding a pointer value to $\loc$. Any subsequent access
to $\loc$ in the continuation must go through dereferencing $\loctmp$, which
resolves to the same $\loc \pointsto v$ fact---there is exactly one owner of
it in the proof context at a time. The borrow's end is not a separate event:
it coincides with whichever later WP rule is the next to consume
$\loc \pointsto v$ again (e.g.\ a subsequent move out of $\loc$, or the drop
of the original binding), at which point the reborrow through $\loctmp$ is
no longer available to the continuation, matching Sec.~\ref{sec:memorymodel}'s
account of borrow-end-via-ownership-transfer.

\subsection{Trusted Computing Base}

\begin{itemize}
  \item \textbf{Trusted:} the Rocq kernel type-checker; \texttt{rustc}'s
    parser, name resolution, and type checker, which determine the THIR AST
    that \texttt{hax}/\texttt{hax2rocq} (Sec.~\ref{sec:motivation}) transpile
    into Rocq; the \texttt{hax}/\texttt{hax2rocq} transpiler itself, which we
    do not independently verify against \texttt{rustc}'s THIR; the pinned
    Rocq, Iris, and \texttt{std++} versions fixed by the project's Nix flake,
    consistent with any foundational Rocq/Iris development.
  \item \textbf{Untrusted (oracle only):} the borrow checker
    (Sec.~\ref{sec:memorymodel}). Its acceptance of a program is never
    consumed as a fact by the dynamic semantics or by any proof; it only
    predicts, ahead of time, that the fractional/affine obligations a
    borrow-checked program generates will in practice be dischargeable.
    A borrow-checker bug can at worst make an actually-safe program harder
    to verify (its obligations may fail to discharge); it cannot make an
    unsound proof go through, since no rule takes borrow-checker output as a
    premise.
\end{itemize}

\subsection{Ownership Threading in the Allocator}

The buddy allocator's methods take \lstmath{\&mut self}, so each call in the
allocator's call graph (Sec.~\ref{sec:casestudy}) both requires and returns
exclusive ownership of the receiver's representation: a caller's precondition
$\mathit{self} \pointsto v$ is consumed by the callee's $\WPfn$, and the
callee's postcondition returns a (possibly updated) $\mathit{self} \pointsto
v'$ to the caller, exactly mirroring the exclusivity \lstmath{\&mut}
enforces syntactically. This threading composes uniformly through the call
graph: an intermediate helper that itself calls further \lstmath{\&mut self}
methods on (a projection of) the same receiver passes the points-to fact
along its own precondition/postcondition pair, so ownership never needs to be
duplicated or reconstructed---each call in the chain hands off the single
outstanding $\mathit{self} \pointsto v$ fact to the next.

The current boundary of what is handled this way is stored references and
reborrows through struct fields: a method that takes \lstmath{\&mut self}
and internally reborrows one of \lstmath{self}'s fields (e.g.\ operating on
\lstmath{self.blocks} or \lstmath{self.heads} individually before returning)
is verified by splitting $\mathit{buddyRep}$ into its field-level
representation predicates ($\mathit{blocksRep}$, $\mathit{headsRep}$,
Sec.~\ref{sec:casestudy}) and recombining them at the end of the method,
following the same ownership-transfer pattern as the worked example above.
References stored inside longer-lived data structures and returned across
multiple call frames beyond this pattern are not exercised by the case
study and are left to future work.
 \fi

\end{document}